\documentclass[twocolumn]{aastex631}
\usepackage{newtxtext,newtxmath}
\usepackage{upgreek}
\usepackage[T1]{fontenc}
\usepackage{graphicx}
\usepackage{MnSymbol}
\usepackage{xspace}
\usepackage{amsmath}	
\usepackage{amssymb}
\usepackage{stackengine}

\newcommand{\dmunit}{$\mathrm{pc~cm}^{-3}$\xspace}

\accepted{October 31, 2023}

\begin{document}

\title{Constraints on the persistent radio source associated with FRB\,20190520B using the European VLBI Network}

\shorttitle{PRS associated with FRB\,20190520B with the EVN}
\shortauthors{Bhandari et al.}
\correspondingauthor{Shivani Bhandari}
\email{bhandari@astron.nl}
\author[0000-0003-3460-506X]{Shivani Bhandari}
\affiliation{ASTRON, Netherlands Institute for Radio Astronomy, Oude Hoogeveensedijk 4, 7991~PD Dwingeloo, The Netherlands}
\affiliation{Joint institute for VLBI ERIC, Oude Hoogeveensedijk 4, 7991~PD Dwingeloo, The Netherlands}
\affiliation{Anton Pannekoek Institute for Astronomy, University of Amsterdam, Science Park 904, 1098~XH Amsterdam, The Netherlands}
\affiliation{CSIRO Space and Astronomy, Australia Telescope National Facility, PO Box 76, Epping, NSW~1710, Australia}

\author[0000-0001-9814-2354]{Benito Marcote}
\affiliation{Joint institute for VLBI ERIC, Oude Hoogeveensedijk 4, 7991~PD Dwingeloo, The Netherlands}

\author[0000-0002-5519-9550]{Navin Sridhar}
\affiliation{Department of Astronomy, Columbia University, New York, NY~10027, USA}
\affiliation{Theoretical High Energy Astrophysics (THEA) Group, Columbia University, New York, NY~10027, USA}
\affiliation{Cahill Center for Astronomy and Astrophysics, California Institute of Technology, Pasadena, CA~91125, USA}

\author[0000-0003-0307-9984]{Tarraneh Eftekhari}
\affiliation{Center for Interdisciplinary Exploration and Research in Astrophysics (CIERA) and Department of Physics and Astronomy,\\Northwestern University, Evanston, IL~60208, USA}

\author[0000-0003-2317-1446]{Jason W.~T. Hessels}
\affiliation{ASTRON, Netherlands Institute for Radio Astronomy, Oude Hoogeveensedijk 4, 7991~PD Dwingeloo, The Netherlands}
\affiliation{Anton Pannekoek Institute for Astronomy, University of Amsterdam, Science Park 904, 1098~XH Amsterdam, The Netherlands}

\author[0000-0002-5794-2360]{Dant\'e M. Hewitt}
\affiliation{Anton Pannekoek Institute for Astronomy, University of Amsterdam, Science Park 904, 1098~XH Amsterdam, The Netherlands}

\author[0000-0001-6664-8668]{Franz Kirsten}
\affiliation{Department of Space, Earth and Environment, Chalmers University of Technology, Onsala Space Observatory, SE-439~92, Onsala, Sweden}

\author[0000-0001-9381-8466]{Omar S. Ould-Boukattine}
\affiliation{ASTRON, Netherlands Institute for Radio Astronomy, Oude Hoogeveensedijk 4, 7991~PD Dwingeloo, The Netherlands}
\affiliation{Anton Pannekoek Institute for Astronomy, University of Amsterdam, Science Park 904, 1098~XH Amsterdam, The Netherlands}

\author[0000-0002-5195-335X]{Zsolt Paragi}
\affiliation{Joint institute for VLBI ERIC, Oude Hoogeveensedijk 4, 7991~PD Dwingeloo, The Netherlands}

\author[0000-0001-6170-2282]{Mark P. Snelders}
\affiliation{ASTRON, Netherlands Institute for Radio Astronomy, Oude Hoogeveensedijk 4, 7991~PD Dwingeloo, The Netherlands}
\affiliation{Anton Pannekoek Institute for Astronomy, University of Amsterdam, Science Park 904, 1098~XH Amsterdam, The Netherlands}

\begin{abstract}
We present very-long-baseline interferometry (VLBI) observations of a continuum radio source potentially associated with the fast radio burst source FRB\,20190520B. Using the European VLBI network (EVN), we find the source to be compact on VLBI scales with an angular size of $<2.3$\,mas ($3\sigma$). This corresponds to a transverse physical size of $<9$\,pc
(at the $z=0.241$ redshift of the host galaxy), confirming it to be an FRB persistent radio source (PRS) like that associated with the first-known repeater FRB~20121102A. The PRS has a flux density of $201 \pm 34\,\mathrm{\upmu Jy}$ at 1.7\,GHz and a spectral radio luminosity of $L_{1.7\,\rm GHz} = (3.0 \pm 0.5) \times 10^{29}\,\mathrm{erg\,s^{-1}\,Hz^{-1}}$ (also similar to the FRB~20121102A PRS). Comparing to previous lower-resolution observations, we find that no flux is resolved out on milliarcsecond scales. We have refined the PRS position, improving its precision by an order of magnitude compared to previous results. We also report the detection of a FRB\,20190520B burst at 1.4\,GHz and find the burst position to be consistent with the PRS position, at $\lesssim20$\,mas.
This strongly supports their direct physical association and the hypothesis that a single central engine powers both the bursts and the PRS. We discuss the model of a magnetar in a wind nebula and present an allowed parameter space for its age and the radius of the putative nebula powering the observed PRS emission. Alternatively, we find that an accretion-powered `hypernebula' model also fits our observational constraints.
\end{abstract}


\section{Introduction} \label{sec:intro}
Fast radio bursts (FRBs) are short-duration bursts ($\upmu$s--ms) of radio waves that typically come from distant extragalactic astronomical sources (3.6\,Mpc to 6.7\,Gpc), predominately in star-forming galaxies \citep{Bhandari+22,Gordon+23}. While most known FRBs appear as one-off events \citep{chimecat}, some are known to repeat \citep{Spitler+16}. One magnetar in our own Milky Way, SGR~1935+2154, produced a bright FRB-like burst \citep{Bochenek+20,2020Natur.587...54C}, suggesting that a fraction of FRBs could have a magnetar origin. However, the diversity of FRB locations and burst properties suggests that a single magnetar origin may be insufficient to explain the observed phenomena in general, and that there may be multiple types of FRB progenitors \citep{Petroff+22}. 

FRBs have spectral luminosities ranging from $10^{27}\text{--}10^{34}$\,erg\,s$^{-1}$\,Hz$^{-1}$ \citep{Petroff+22} and are mostly characterized by
high dispersion measure (DM) relative to the expectations of Galactic electron density models \citep{ne2001, YMW16}. While the exact nature and origins of FRBs are still a subject of ongoing research and debate, one intriguing aspect that has emerged is the apparent association of a persistent radio source (PRS) with FRB\,20121102A and potentially FRB\,20190520B \citep{VLAlocalisation,Niu+22}. PRSs are defined as being continuum radio sources that are distinct from radio emission caused by ongoing star formation in the host galaxy \citep{Bhandari+20,Nimmo+22b,Dong+23}. They are too luminous \citep[L$_{\rm PRS} > 10^{29}$\,erg\,s$^{-1}$\,Hz$^{-1}$;][]{Law+22} and too compact \citep[$< 1$\,pc;][]{Marcote17} to plausibly be related to star formation. Rather, they may be powered by the same central engine that creates the bursts themselves. 

FRB\,20121102A and FRB\,20190520B are `twin' sources that are both active repeaters in low-mass host galaxies, and are embedded in dynamic magneto-ionic environments \citep{MichilliRM,Anna-Thomas+22}. The FRB\,20121102A bursting source is co-located with a PRS. The compactness of the radio source associated with FRB\,20190520B, on the other hand, is not yet directly established. The FRB\,20121102A-associated PRS and FRB\,20190520B-associated PRS candidate have a flat spectrum from 1.6--10\,GHz with a spectral index of $\alpha = -0.27 \pm 0.24$ and $ -0.41 \pm 0.04$, respectively \citep{Marcote17, Niu+22}, where S$_{\nu} \propto \nu^{\alpha}$. The PRS spectrum for FRB\,20121102A also remains optically thin down to 400\,MHz with a flat spectral index \citep{Resmi+21}. However, it becomes steeper ($\alpha \sim -1.2$) at frequencies $>10$\,GHz \citep{VLAlocalisation}.
Both PRSs have a flux density in the range of $180$--$200$\,$\upmu$Jy at 3\,GHz and spectral radio luminosities of the order of $10^{29}$\,erg\,s$^{-1}$\,Hz$^{-1}$ \citep{VLAlocalisation,Niu+22,Zhang+23}. We highlight that, given the existing data, we cannot conclude if the luminosity similarity is intrinsic, coincidental, or due to observational biases.

The flux density of FRB\,20121102A's PRS is observed to vary by $\sim 10\%$ at 3\,GHz on day timescales, which is consistent with refractive scintillation in the Milky Way \citep{VLAlocalisation, Waxman+17}. There is apparently no link between flux density fluctuations of the PRS and burst activity. \citet{Chen+22} monitored the PRS at $12-26$\,GHz and found the level of radio flux variability to be lower than the expectations from scintillation given the source's compact size, ruling out active galactic nuclei (AGNe) as possible model for FRB\,20121102A's PRS. In a recent study, \citet{Rhodes+23} found a $\sim30$\% change in the flux density of the PRS at 1.3\,GHz over three years which they argue to be more likely intrinsic to the source than due to scintillation. 
However, more measurements of the PRS over a range of timescales are required to robustly rule out the possibility of scintillation. In the case of FRB\,20190520B's PRS, \citet{Zhang+23} reported a marginal $3.2\sigma$ decrease in the flux density at 3\,GHz over a timescale of a year. This decrease could either be intrinsic or due to scintillation, which could limit the size of the potential variable component of the radio source to sub-parsec level.  

Very-long-baseline interferometry (VLBI) observations of repeating FRBs with accompanying PRSs can provide insights into the nature of the burst source and the PRS, as well as their potential relation. For instance, observations with the European VLBI Network (EVN) have strongly constrained the size of FRB\,20121102A's associated PRS to be $<0.7$\,pc ($1\sigma$) with a projected linear separation of $<40$\,pc ($1\sigma$) from the location of the burst source \citep{Marcote17}. Moreover, a steady flux density of the PRS over a year after detection disfavored a young supernova scenario \citep{Plavin+22}. Finally, the level of polarization is observed to differ significantly between the burst and PRS emission. While the burst from FRB\,20121102A is $>90\%$ polarized at $4.8\,\mathrm{GHz}$ \citep{MichilliRM}, the PRS is unpolarized with conservative upper limit of $<25$\%
at $4.8\,\mathrm{GHz}$ \citep{Plavin+22}. This ruled out the possibility that the bursts and PRS are of the same nature, i.e., persistent emission is not driven by a buildup of regular, low-level burst activity from the FRB\,20121102A source \citep[as also shown previously in ][]{Gourdji+19}.

FRB\,20121102A's dwarf host galaxy, association with a PRS, and the high rotation measure (RM) of its bursts   \citep{MichilliRM} led to a concordance picture of an FRB source as a flaring magnetar embedded in a magnetized wind nebula, where the putative young (30--100\,yr old) magnetar is formed as a remnant from a super luminous supernova (SLSN) or long gamma-ray burst \citep[LGRB;][]{Margalit+18}. In this scenario, the persistent emission is powered by relativistic electrons heated at the termination shock of the magnetar wind, while the RM originates from non-relativistic electrons injected earlier in the nebula’s evolution and cooled through expansion and radiative losses. As mentioned previously, VLBI observations have provided tight constraints on the size of the radio nebula \citep{Marcote17}, showing that it must be at least twice as small as the Crab nebula. Such a model is able to explain the observed size and luminosity of the PRS, as well as the large and decreasing RM of the bursts \citep{Hilmarsson+21}. \citet{Yang+20} suggest that the PRS luminosity might be related to the high RM of the FRB source. Other models include a `cosmic comb' in which an astrophysical gas flow (stream) interacts with the magnetosphere of a foreground neutron star to produce an FRB \citep{Zhang+18}. Finally, a model involving a `hypernebula' is proposed, where an accreting compact object is able to produce FRBs along a jet cone, and the surrounding turbulent baryon rich outflow from the accretion disk is responsible for the persistent radio emission, and accounts for the overall decreasing and fluctuating RM \citep{Sridhar&Metzger22}. The baryons accelerated at the jet termination shock of these potential PRS sources could also be sources of persistent high-energy neutrinos \citep{Sridhar+23b}.

FRB\,20190520B, a repeating FRB discovered at a DM of $1205\pm 4$\,pc\,cm$^{-3}$ by the FAST telescope, is observed to share similar burst and host properties to FRB\,20121102A. It originates in an extreme magneto-ionic environment \citep{Anna-Thomas+22} in a dwarf galaxy located at a redshift $z = 0.241$. For the given redshift, FRB\,20190520B has an estimated host galaxy contribution to the DM of $903_{-111}^{+72}$\,pc\,cm$^{-3}$. This contribution is unlikely to be produced by the interstellar medium of the host galaxy, but rather more plausibly originates from the local environment of the source which may also be linked to the co-located radio source. This host DM is a factor of $\sim 5$ larger than what is observed for typical FRB host galaxies \citep{James+22b} and a factor of a few beyond what is estimated for FRB\,20121102A \citep{Tendulkar17}. Using Karl G.\ Jansky Very Large Array (VLA) observations, the size of the apparent PRS was confined to $<1.4$\,kpc \citep{Niu+22}. Recently, based on equipartition and self-absorption assumptions, a lower size limit of $\geq 0.22$\,pc was placed using the radio spectrum and the integrated radio luminosity in the 1--12\,GHz range \citep{Zhang+23}.

In this Letter, we present EVN observations of the continuum radio source associated with FRB\,20190520B, directly showing that it is a compact (parsec-scale) PRS and improving the constraints on its transverse physical size and position by over an order-of-magnitude. We also demonstrate that the burst source and PRS are likely to be strictly co-located, and thus most likely powered by the same central engine. After FRB\,20121102A, FRB\,20190520B is now only the second FRB source to demonstrate these characteristics, which strongly inform potential models for the source's nature. In \S\ref{sec:section2}, we describe our observations and data analysis. We present our results in \S\ref{sec:section3} and discuss the implication of our results in \S\ref{sec:section4}. Finally, we present our conclusions in \S\ref{section:conclusions}.

Throughout the paper we employ a standard cosmology of $H_{0}=~67.7$~km~s$^{-1}$~Mpc$^{-1}$, $\Omega_{\rm M}=~0.31$, $\Omega_{\rm vac}=~0.69$ \citep{Planck18}.

\section{Observations and data analysis}  \label{sec:section2}
We observed the field of FRB\,20190520B using the EVN at a central frequency of 1.7\,GHz, with the total bandwidth divided into $4 \times 32$-MHz subbands (project code EM161; PI: Marcote). The observations were performed at two epochs on 2022 February 26 and 27 (EM161A and EM161B, respectively) both from 01:30--08:30~UT. Eleven EVN dishes --- namely, Westerbork single dish RT1 (Wb), Effelsberg (Ef), Medicina (Mc), Noto (Nt), Onsala (O8), Tianma (T6), Urumqi (Ur), Toru\'n (Tr), Hartebeesthoek (Hh), Irbene (Ir), and Jodrell Bank Mark II (Jb) --- participated in these sessions.

We also observed this field as part of our ongoing FRB VLBI localization project, PRECISE (Pinpointing REpeating ChIme Sources with Evn dishes) under the project code PR236A (PI: Kirsten). Only Ef, Tr, Nt and Wb participated in this observation conducted on 2022 August 17 at a central frequency of 1.4\,GHz. While the PR236A observation lasted from 19:30--05:30~UT, only the first two hours were focused on the field of FRB\,20190520B. These data were correlated at the Joint Institute for VLBI ERIC (JIVE) under the project code EK051C (PI: Kirsten) with the total bandwidth divided into $8 \times 32$-MHz subbands. However, we note that Nt and Wb only observed in 6 and 4 of these subbands, respectively.

We correlated the EM161A/B data for the field of FRB\,20190520B at a position consistent with the published VLA position, which has an uncertainty of 100\,mas and 50\,mas in $\alpha$ and $\delta$, respectively \citep{Niu+22}. We interleaved 4.5-min scans on the target source with 1.2-min scans on the phase calibrator source, J1605$-$1139, located $0.87^\circ$ away from FRB~20190520B. The sources J1550+0527 and J2253+1608 were also observed as fringe-finder and bandpass calibrators for EM161A/B, and J1642$-$0621 and J1927+6117 for EK051C. We also observed J1603$-$1007 as a check source to test our calibration and phase-referencing technique as well as the final astrometric accuracy.

\subsection{EVN interferometric data}   

The continuum interferometric data were correlated using the software correlator SFXC \citep{Keimpema+15} at JIVE, with an integration time of 2\,s and 64 channels per each 32-MHz subband. The data were calibrated following standard procedures in AIPS \citep{AIPS}, CASA \citep{CASA,Bemmel_2022}, and {\sc Difmap} \citep{DIFMAP}. In order to verify our results, we followed two parallel data reduction procedures in AIPS and CASA. The correlated visibilities in FITS-IDI format were loaded into AIPS using the \texttt{FITLD} task. The a-priori amplitude calibration (performed using the known gain curves and system temperature measurements recorded on each station during the observation) and a-priori flagging table were applied from the tables generated by the EVN AIPS pipeline.

We also used \texttt{VLBATECR} in AIPS to correct for ionospheric dispersive delays. This task downloads the IONEX files provided by the Jet Propulsion Laboratory (JPL) containing the total electron content (TEC) maps of the ionosphere at the time of the observation and calibrates the data based on them for the different antenna sites. 

For the data reduction in CASA, we imported the data in UVFITS format with the aforementioned calibration using the task \texttt{importuvfits}. This task converted such data to a CASA measurement set (MS). The data were inspected using \texttt{plotms} and 10\% of the edge channels per spectral window and the first 5\,s of each scan were flagged using the \texttt{flagdata} task. 

Next, we performed delay and phase calibration using the task \texttt{fringefit} \citep{Bemmel_2022}. This was accomplished in two steps: 1) single-band delay correction, which corrects for instrumental effects between subbands; 
2) multi-band delay correction, which performs global fringe fitting across all data on the calibrator sources. We used the best fringe-finder source scan for single band delay correction. Within the scan a solution is determined for each spectral window. For multi-band delay, we correct for phases as a function of time and frequency for all the phase-referencing and fringe-finder source scans. The final step in the calibration is the bandpass correction, which was performed using CASA’s \texttt{bandpass} task. We used the best fringe-finder scan data for bandpass calibration. 
The single-band delay and multi-band delay calibration tables, along with the bandpass calibration table, were applied to the measurement set using the task \texttt{applycal}. The phase calibrator, target and check source were averaged in frequency and split into single source files using the task \texttt{split}. We first imaged and self-calibrated the phase-referencing source using tasks \texttt{tclean} and \texttt{gaincal} to obtain the best possible model of the source. This model allowed us to improve the phases and amplitudes of the different antennas, which in return led to an improved calibration of the check source and target. The respective MS data were then converted to UVFITS format using \texttt{exportuvfits}.

In the case of the second approach, with a data reduction fully within AIPS, we repeated the same steps but using the standard AIPS tasks with equivalent parameters \citep[as done in, e.g,][]{Marcote17, Marcote+20, Nimmo+22b}. In this case, we used {\sc Difmap} for imaging and self-calibration of the phase-referencing source and the resulting model was imported into AIPS for improving the calibration of the check source and target. 

Finally, we imaged both the target (FRB\,20190520B) and the check source (J1603$-$1007) using {\sc Difmap}. We combined the data from two epochs (EM161A and EM161B) to achieve better sensitivity and image fidelity. We note that the Tr dish was observed to show systematic phase variations with time and therefore its data were flagged during imaging. We used a cell size of 0.2\,mas and natural weighting to image the target source (referred to as `image I'). Additionally, given the low elevation of the source during our observations (highest elevation of $20$--$40^\circ$ for most European antennas), 
another set of images were obtained using the data for which the source had an elevation greater than $20^\circ$ ($\sim75\%$ of the total data; referred to as `image II'). In \S\ref{subsec:PRS}, we will use images I and II to characterize the astrometry of the source. 
\\
\\
\subsection{Single-pulse search}

The baseband data recorded at Ef for the three aforementioned observations (EM161A, EM161B, and PR236A/ EK051C) were searched for bursts from FRB\,20190520B using the PRECISE data analysis pipeline\footnote{\url{https://github.com/pharaofranz/frb-baseband}}. The baseband data were channelized with a time resolution of 128\,$\upmu$s and a frequency resolution of 125\,kHz using \texttt{digifil} \citep{van_Straten+11}. The resulting total-intensity filterbank data products \citep{Sigproc} were searched for single pulses with \texttt{Heimdall}\footnote{\url{https://sourceforge.net/projects/heimdall-astro/}} using a detection threshold of $7\sigma$ and a DM range $1202\pm50$~\dmunit. Burst candidates were classified with \texttt{FETCH} \citep{Agarwal:2020} using models A and H with a 50\% probability threshold to separate likely astrophysical events from false positives. The final set of candidates was inspected by eye to confirm their astrophysical nature. The single-pulse search pipeline is explained in greater detail by \cite{Kirsten+21}. 
\\
\\
\begin{figure}
    \includegraphics[width=0.48\textwidth,trim={0.2cm 0.2cm 0 0},clip]{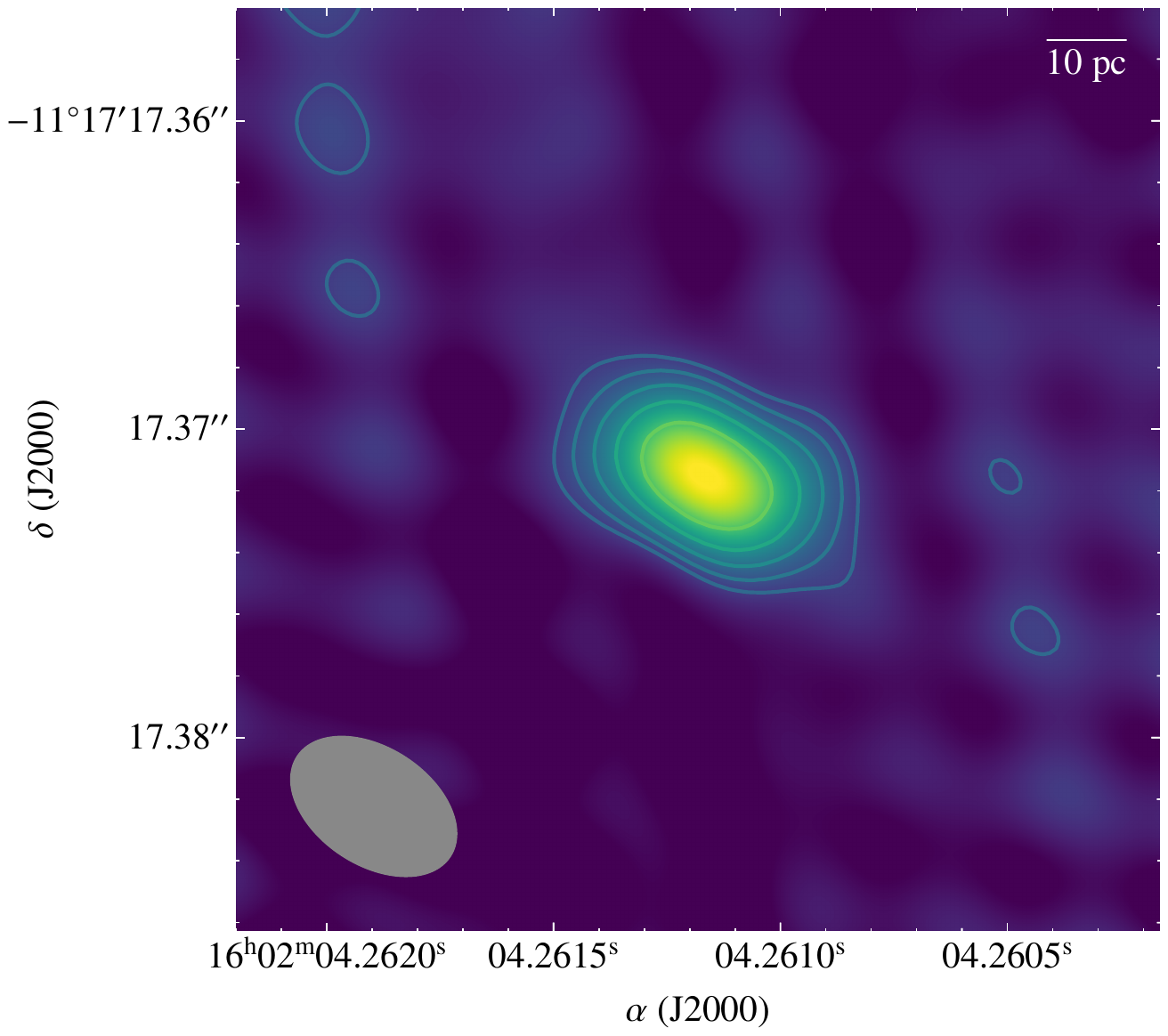} \\
    \caption{EVN image of the PRS associated with FRB~20190520B, as seen in the combined EM161A/B observations. The source is compact on milliarcsecond scales: $<2.3~\mathrm{mas}$ at $3\sigma$ level. This corresponds to a  $<9~\mathrm{pc}$ transverse extent, given the known redshift of the host galaxy. A small bar in the top-right of the image shows a representative 10-pc transverse extent, for scale. Contour levels start at two times the rms noise level of $16~\mathrm{\upmu Jy\ beam^{-1}}$ and increase by factors of $\sqrt{2}$. The synthesized beam is represented by the gray ellipse at the bottom left corner; it has a size of $3.8 \times 5.9~\mathrm{mas^2}$ and a position angle of $57^\circ$.}
    \label{fig:em161}
\end{figure}

\begin{figure}
\centering
\includegraphics[width=0.3\textwidth,trim={1.6cm 0cm 1.6cm 0cm}]{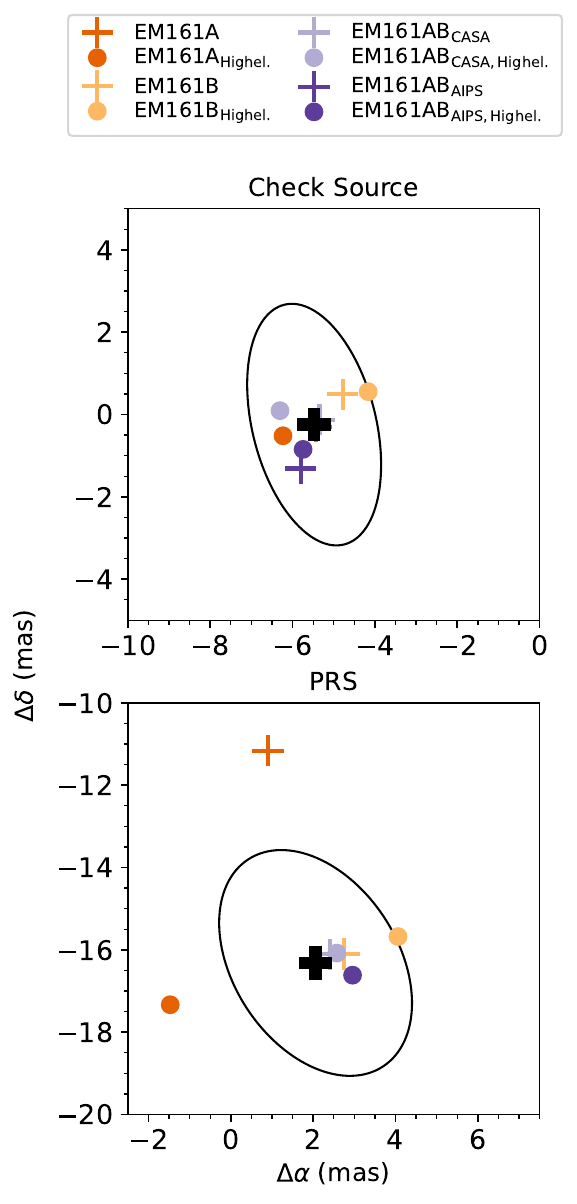}
\caption{{\it Top panel}: Comparison of the position of the check source (J1603$-$1007) measured from EM161A (orange) and EM161B (light orange) for all data (plus markers) and only high-elevation data (circles), all relative to the Astrogeo position at 3\,GHz. Also plotted are the positional offsets measured from CASA-analysis (light purple) and AIPS-analysis (violet) for the combined-epoch EM161A/B data. The filled black plus shows the average position derived from individual measurements at different epochs, using different analysis software, and different data selections. The gray ellipse shows the size of the synthesized beam for the combined EM161A/B data, centered at the average position. The individual measurements fall within the FWHM of the synthesised beam.
{\it Bottom panel}: Measured positional offsets for the PRS relative to the centroid of the published VLA position \citep{Niu+22}. We note that the uncertainty on this position is 100\,mas and 50\,mas in $\alpha$ and $\delta$ respectively, which is larger than the extent of the image. Here the individual measurements show a scatter of a few mas with respect to the average PRS position shown by the dark black cross.}
\label{fig:astrometry}
\end{figure}

\section{Results}  \label{sec:section3}
\subsection{Persistent radio source} \label{subsec:PRS}

A compact PRS is detected in the EVN data. Figure~\ref{fig:em161} shows the continuum image of the region at 1.7\,GHz as seen with a synthesized beam size (full width at half maximum, FWHM) of $3.8 \times 5.9\,\mathrm{mas^2}$ and a root-mean-square (rms) noise level of $16\,\upmu\mathrm{Jy\,beam^{-1}}$. The source detected in the combined EM161A/B dataset has a peak brightness of $186 \pm 32\,{\rm \upmu Jy\,beam^{-1}}$ and an integrated flux density of $201 \pm 34\,\mathrm{\upmu Jy}$.
This flux density translates to a spectral radio luminosity of $L_{1.7\,\rm GHz} = (3.0 \pm 0.5) \times 10^{29}\,\mathrm{erg\,s^{-1}\,Hz^{-1}}$ at the known luminosity distance of 1.25\,Gpc. We find consistent flux density values between the EM161A ($197\pm 40\,\mathrm{\upmu Jy}$) and EM161B ($210 \pm 40\,\mathrm{\upmu Jy}$) observations, which are separated by a day. We note that the uncertainty on the flux density is the quadrature sum of the rms noise and 15\% of the absolute flux density error, which is typical for VLBI observations. 

We constrained the apparent angular sizes of the observed sources by $\chi^2$ fitting of a circular Gaussian model in the $uv$-plane.
This is more robust than fitting an elliptical beam since the solution, to some extent at least, will not be degenerate with the beam's ellipticity. We obtained an apparent angular source size of $1.4 \pm 0.3\,\mathrm{mas}$ for FRB\,20190520B's associated PRS, and sizes of $\approx 2.0\,\mathrm{mas}$ and $\approx 5.9\,\mathrm{mas}$ for the core components of the phase calibrator (J1605$-$1139) and check source (J1603$-$1007), respectively. A comparatively larger source size for the check source hints at extended emission.

\subsubsection{Astrometry}
We investigated the astrometry by applying the derived calibration solutions to the check source followed by imaging. Our comparison of the position of the check source
measured from images created with all data (image I) and solely high-elevation data (image II) at both epochs (EM161A and EM161B) individually, as well as the combined-epoch (EM161A/B) data set are presented in the top panel of Figure~\ref{fig:astrometry}.
We also compared the combined-epoch position from CASA data analysis to that of AIPS. 
We find the position of the core emission of the check source in our images to have an offset of $\Delta\,\alpha\ = -5.5$\,mas and $\Delta\,\delta\ = -0.2$\,mas compared to the position derived from the 3-GHz image in the Astrogeo catalog\footnote{\url{http://astrogeo.org/sol/rfc/rfc_2023a/rfc_2023a_cat.html}}. Because of the different central observing frequencies between our data (1.7\,GHz) and the Astrogeo (3\,GHz) reference position of the check source, we cannot rule out the possibility that these offsets are caused (in part) by core-shift --- i.e., a change in the apparent position of the jet base (core) in radio-loud AGN with frequency, due to synchrotron self-absorption. As a result, we chose not to apply these offsets in calculating the PRS location from our EVN data but conservatively include them in our systematic uncertainty calculations. 

We further compare the position of the PRS measured at different epochs using different analysis software and data selections. These comparisons are presented in the bottom panel of Figure~\ref{fig:astrometry}. We note that the EVN-PRECISE data (EK051C) are not included in this comparison or the combined data set because of their poor $uv$-coverage. For data reduced in CASA, we find the PRS position from the combined EM161A/B data to be consistent between images I and II. This position is also consistent with the image II position (within $<1$\,mas) measured in the AIPS-reduced data. However, we do find an offset of $\sim 5$~mas for image I between the CASA and AIPS reductions. We note that comparable flagging was performed in both AIPS and CASA, implying that the positional difference is unlikely to be due to flagging. Moreover, we find the standard deviation of the offsets derived from EM161A and EM161B considering all data and high-elevation-only data ($\sigma_1$), and also compute the standard deviation from the combined epoch data analyzed in AIPS/CASA with all and high-elevation-only data ($\sigma_2$). We find $\sigma_1 > \sigma_2$, and thus we conservatively use $\sigma_1(\alpha,\delta)=~(2.1, 2.3)$\,mas as a measurement of the scatter in the positional offsets. 

We average the above positional offsets and measure the final position of the FRB\,20190520B PRS within the International Celestial Reference Frame (ICRF):
\\
\indent$\alpha (\text{J}2000) = 16^{\rm h}02^{\rm m}04.2611^{\rm s} \pm 6.5\,\rm{mas}$, \\
\indent$\delta (\text{J}2000) = -11^\circ17^\prime17.366^{\prime\prime} \pm 3.6~\mathrm{mas}$.\\ 
The quoted uncertainties take into account the statistical uncertainties on the measured position derived from the shape and size of the synthesized beam normalized by S/N ($\Delta \alpha = 0.6\,\rm mas~, \Delta \delta = 0.4$\,mas); the systematic uncertainties of the absolute position of the phase calibrator (J1605$-$1139; $\pm0.1$\,mas) and check source (J1603$-$1007; $\pm0.3$\,mas) within the ICRF; the uncertainties expected from the phase-referencing technique due to the separation between the phase calibrator source and the target source \citep[$\pm 2.5$\,mas;][]{Kirsten+15}; an estimate of the frequency-dependent shift in the phase calibrator position from the ICRF \citep[conservatively $\pm1$\,mas;][]{Plavin+19}; the check source positional offset of $\Delta\,\alpha\ = -5.5$\,mas and $\Delta\,\delta\ = -0.2$\,mas; and a scatter in the PRS position, $\sigma_1(\alpha,\delta)=~(2.1, 2.3)$\,mas derived from the above. Since the PRS position is only $\sim$20\,mas offset from the phase center, no re-correlation of the EM161 data was necessary.

The centroid of our PRS position has an angular offset of 16\,mas with respect to the centroid of the VLA position \citep{Niu+22}, however our PRS position is fully consistent within the uncertainties: $\pm100$\,mas, $\pm50$\,mas in ($\alpha,\delta$) of the VLA position. Our EVN measurements improve the precision on the PRS position by more than an order of magnitude.

\begin{figure}
    \centering
    \includegraphics[width=0.45\textwidth]{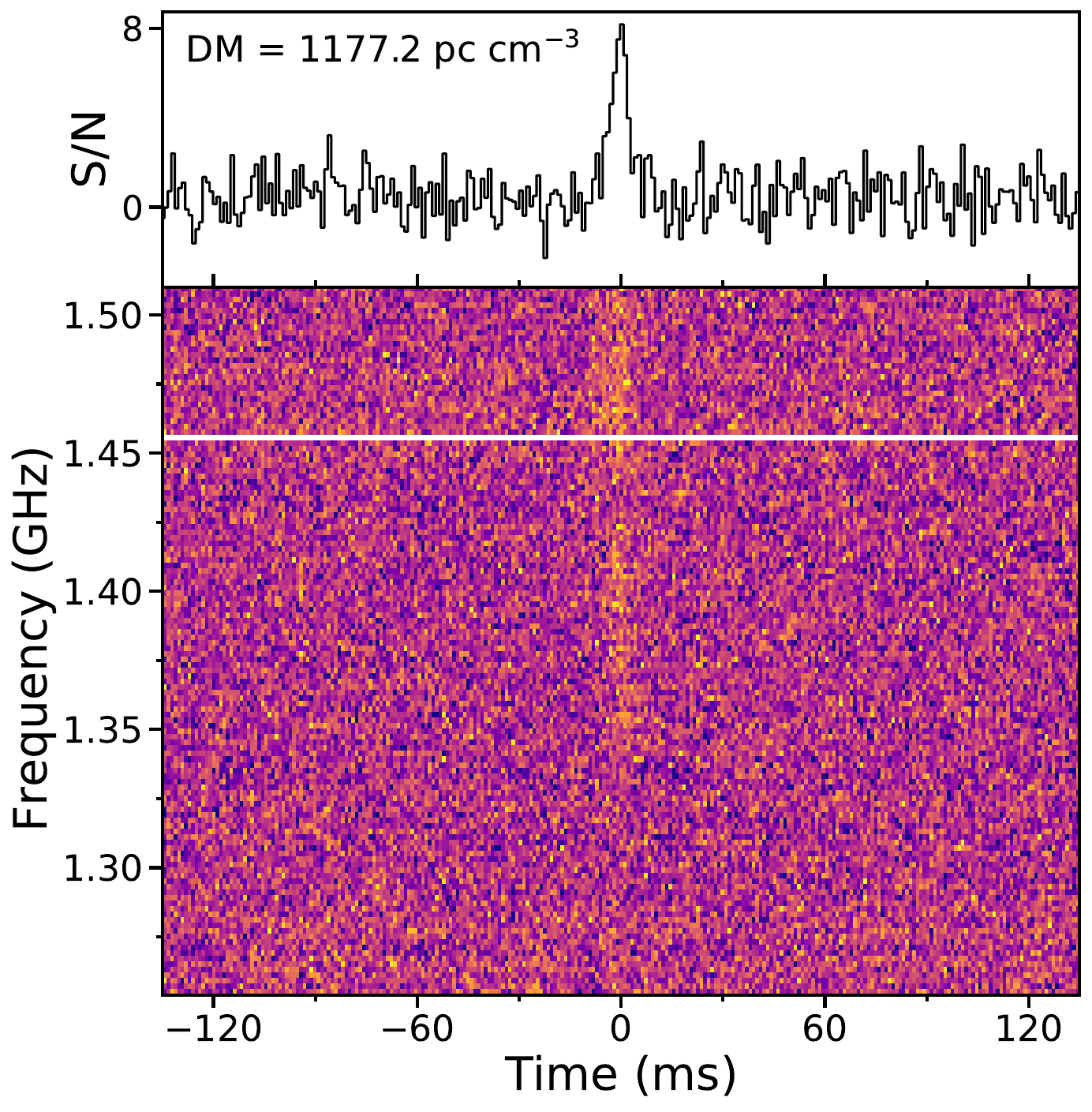}
    \caption{The burst profile (top panel) and dynamic spectrum (bottom panel) of the only burst detected from FRB\,20190520B in our observations (in EK051C). The time and frequency resolution are 1.024\,ms and 2\,MHz, respectively. The data has been coherently dedispersed using SFXC at the \texttt{Heimdall} reported DM of 1177.2\,pc\,cm$^{-3}$.}
    \label{fig:pulse}
\end{figure}

\subsection{Burst discovery} \label{subsec:burst_discovery}

We detected a single burst from FRB\,20190520B in our EVN-PRECISE observations (EK051C) at a center frequency of 1.4\,GHz. The burst has a time of arrival (TOA) of MJD\,$59808.83055426728$ and was detected with a signal-to-noise (S/N) of $\sim 8$ at a DM of $1177.2$\,pc\,cm$^{-3}$ (integrated over the whole observing band and using a time resolution of $1.024$\,ms). The FRB has a fluence of $1.2 \pm 0.2$\,Jy\,ms and a duration of $7.8 \pm 0.1$\,ms. No additional bursts were detected in any of the other observations (EM161A and EM161B). Figure~\ref{fig:pulse} shows the coherently de-dispersed pulse profile and dynamic spectrum of the burst at a time and frequency resolution of $1.024$\,ms and $2$\,MHz, respectively. The data was coherently de-dispersed using SFXC. The quoted TOA has been corrected\footnote{\url{https://github.com/MSnelders/FRB-Burst-TOA}} to the Solar System Barycentre to infinite frequency assuming a dispersion measure of $1177.2$\,pc\,cm$^{-3}$, reference frequency of $1494$\,MHz and a dispersion constant of $1/(2.41 \times 10^{-4})$\,MHz$^{2}$\,pc$^{-1}$\,cm$^{3}$\,s. The quoted time is in Barycentric Dynamical Time (TBD).

Using the burst TOA in the Ef data, a second correlation was performed on the full EVN data containing the burst, where the gate width used for correlation was determined by eye to maximise the S/N. We used the EVN-derived PRS position from EM161A/B as the phase center for EK051C continuum and burst data. The correlated bin data containing the burst were converted to FITS-IDI format and appended for gain curve and system measurement corrections. We used the CASA task \texttt{importfitsidi} to convert the FITS-IDI file to MS format. The calibration tables obtained during the continuum data reduction of EK051C were also applied to the correlated FRB data, which was later imaged using the task \texttt{tclean} for burst localization.

\begin{figure*}
    \centering
    \includegraphics[width=\textwidth]{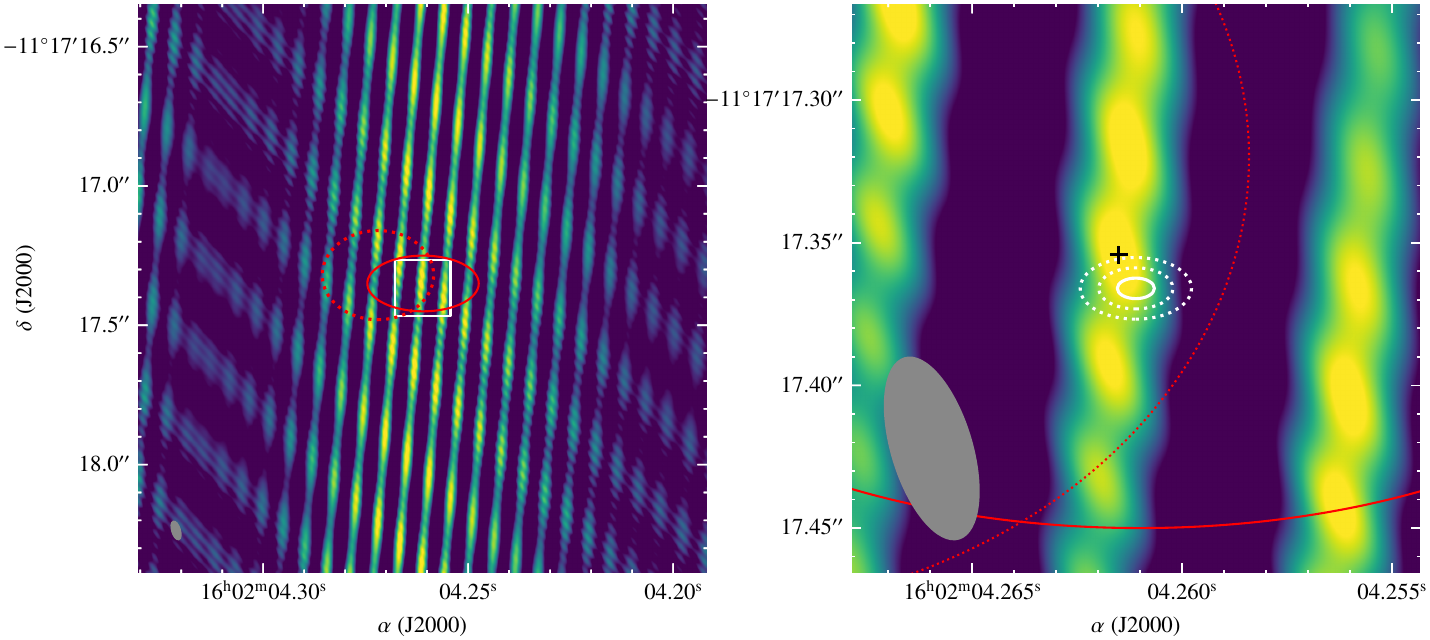} \\
    \includegraphics[width=0.5\textwidth]{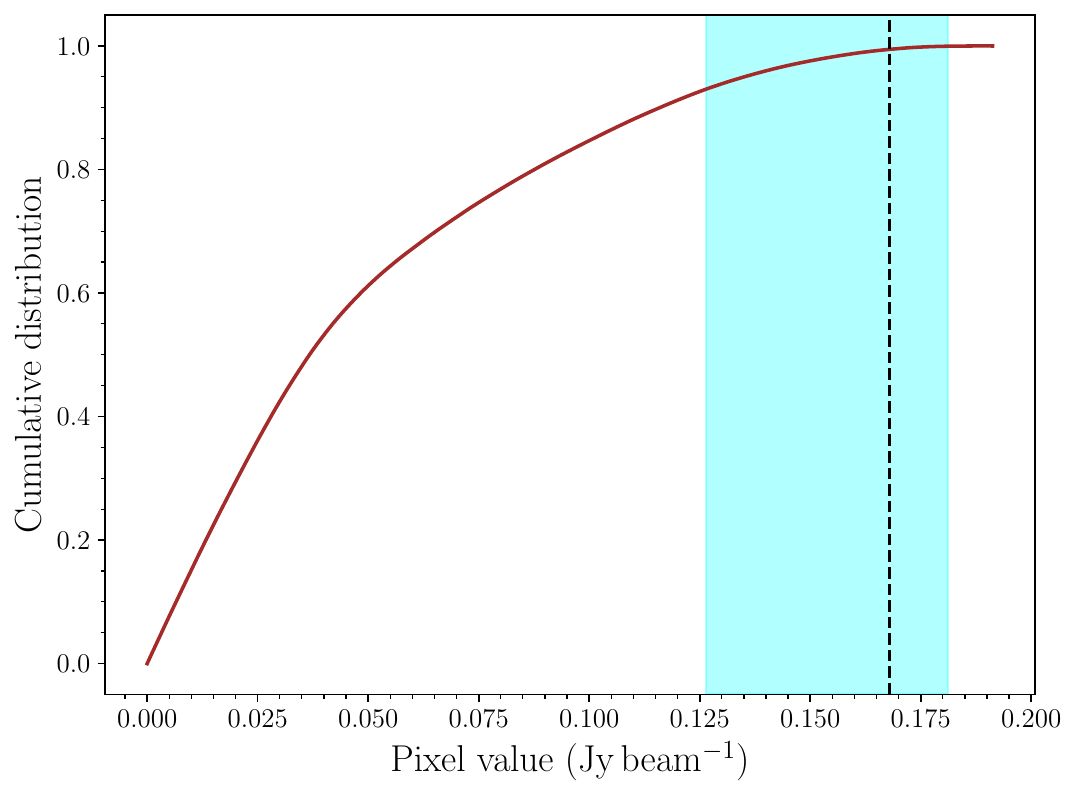} 
    \caption{Direct convolution (the so-called `dirty map') of the interferometric data from the single FRB\,20190520B burst detected in our EVN-PRECISE observations (EK051C). The synthesized beam, with size $62 \times 18\,\mathrm{mas}^2$ and position angle $17^\circ$, is shown as the gray ellipse at the bottom-left corner of each of the two top panels. Also in both top panels, the red solid and dotted ellipses represent the position and $2\sigma$ uncertainty of the PRS and bursting source, respectively, as measured by \citet{Niu+22} using the VLA. {\em Left top}: A $\sim2\times2^{\prime\prime}$ image of the field. The fringe pattern with strong peaks is due to the limited $uv$-coverage resulting from only four available antennas and a few milliseconds of integration (meaning negligible Earth rotation during this time). {\em Right top}: Zoom-in on the white square shown in the left panel. The cross represents the centroid of the strongest peak in the whole $\sim2\times2^{\prime\prime}$ field. The EVN position of the PRS along with the $1\sigma$ (solid), $2\sigma$, and $3\sigma$ (dotted) uncertainties are represented by white ellipses. The PRS position is $\sim 20$\,mas offset from the centroid of the brightest and closest peak.
    {\em Bottom:} A cumulative distribution function of the pixel values in the $\sim2\times2^{\prime\prime}$ field. The black dotted line shows the pixel value at the position of the PRS centroid, while the shaded region represents pixel values within $1\sigma$ of the nominal PRS position. The data are consistent with the PRS and burst source being positionally coincident to within $\lesssim20$\,mas (i.e., to within a transverse distance of $\lesssim80$\,pc).}
    \label{fig:ek051c}
\end{figure*}

Similar to the continuum data analysis, we also conducted a parallel data reduction in {\tt AIPS}. The calibration tables obtained from the continuum data were copied and applied to the burst data, which was finally imaged in {\tt Difmap}.
Figure~\ref{fig:ek051c} shows the direct convolution of the data (the so-called `dirty image') containing the burst. 
Due to the limited $uv$-coverage of this observation at the time of the burst, we cannot independently measure the position of the burst emission with the same accuracy as for the PRS. However, by combining with the PRS information, we can ascertain if the two are at least consistent with being at the same position at the mas-level. 

As can be seen in the left panel of Figure~\ref{fig:ek051c}, the burst signal is spread over several high-amplitude peaks. Nonetheless, the strongest peak in the whole $\sim2\times2^{\prime\prime}$ image is only $\sim 20~\mathrm{mas}$ away from the measured PRS centroid, suggesting that both sources (PRS and burst source) may be coincident (see the right panel of Figure~\ref{fig:ek051c}). To further quantify this, we present a cumulative distribution function (CDF) of the pixel values in the full burst image in the bottom panel of Figure~\ref{fig:ek051c}. The PRS centroid pixel value and those within $1\sigma$ of the PRS position are also marked. This shows that the burst image is consistent with the hypothesis that the PRS and burst source are strictly coincident on mas scales: less than 1\% of the burst image pixels have a larger flux density than what is found at the nominal centroid of the PRS. This is unlikely ($p$-value $< 0.05$) to occur by chance, suggesting that the brightest peak in the burst image indeed represents the burst source position.


\begin{figure}
\includegraphics[width=0.47\textwidth,clip,trim={0cm 0cm 0cm 0cm}]{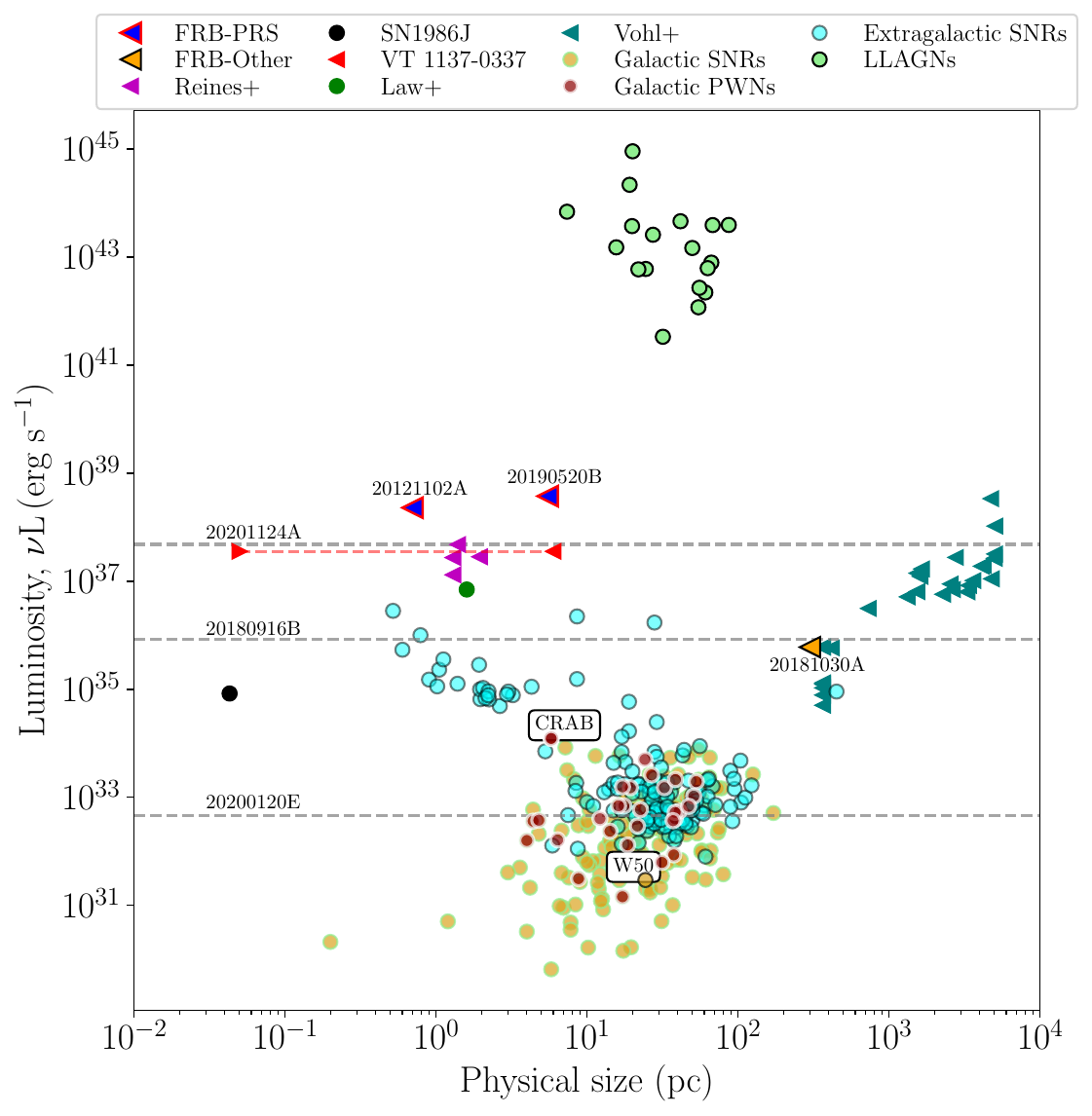}
\caption{Radio luminosities (at 1\,GHz) and physical sizes of multiple types of compact radio sources. These include measurement and constraints for the confirmed FRB PRSs (blue); the radio source potentially associated with FRB\,20181030A \citep[orange;][]{Bhardwaj+21}; 
the compact radio component of SNR SN1986J (black); the orphan GRB afterglow, FIRST~J141918.9+394036 (green); and wandering black hole candidates (magenta) obtained at the highest-possible angular resolutions \citep{Marcote17,SN1986J, Marcote2019,Sargent2022}. Teal points represent the PRS-like candidates identified in nearby dwarf galaxies by \citet{Vohl2023}. Red data points show the inferred lower and upper limits on the size of the PWN candidate VT~1137$-$0337 \citep{Dong2023}. Yellow and dark red points present measurements from Galactic SNRs and PWNe,
respectively \citep{Ranasinghe+2023, Green2019}, whereas cyan points represent SNRs in nearby galaxies \citep{eSN}. The high luminosity space is occupied by light green points showing low-luminosity AGNe \citep{Radcliffe+18}. Circles indicate a size measurement while $1\sigma$ upper limits on size are shown as triangles. Overplotted dashed lines show the $3\sigma$ upper limits on radio luminosity for a sample of localized FRBs presented in Table~\ref{luminosities}. }
\label{fig:LumSize}
\end{figure}

\section{Discussion}  \label{sec:section4}

\subsection{PRS luminosity and size constraints} \label{sec:dis-lum-size}

The spectral radio luminosity of the PRS associated with FRB\,20190520B, determined from our EVN observations, $L_{1.7\,\rm GHz} = (3.0 \pm 0.5)~\times~10^{29}\,\mathrm{\,erg\,s^{-1}\,Hz^{-1}}$ (and its flux density, $S_{1.7 \rm GHz} = 201 \pm 34\,\mathrm{\upmu Jy}$) agrees well with the results previously published by \citet{Niu+22}. 
 
The agreement between these values implies that there is no significant flux density resolved out on VLBI scales ($150\times$ higher resolution than the VLA), thus no significant emission on scales of $\gtrsim 10\,\mathrm{pc}$. The source responsible for the continuum emission is thus compact on those scales. These results likely suggest that the emission of the PRS at 1.7\,GHz remains stable within uncertainties  over a two-year timescale.  Table~\ref{luminosities} compares the PRS luminosities (or their $3\sigma$ upper limits) for various well-localized repeating FRBs. These are also presented in Figure~\ref{fig:LumSize}. FRB\,20121102A and FRB\,20190520B are the only FRBs associated with a PRS and have consistent luminosities that are much greater than the upper-limits for the other FRBs. These luminosities, together with other observed burst properties \citep{MichilliRM, Hilmarsson+21,Anna-Thomas+22}, suggest that these two FRBs may be particularly young and active sources surrounded by dense and dynamic magnetized plasma. For instance, in the case of FRB\,20121102A, for a scenario where the magnetic energy of the magnetar inflates a synchrotron nebula behind the expanding supernova ejecta, the age of the magnetar, which is inversely proportional to the luminosity (see Equation\,\ref{eq:tage}), is estimated to be 10--100\,yr \citep{Margalit+18}.

\begin{table*}
\centering
\begin{tabular}{lrll}
\hline
\hline
FRB name & Luminosity\phantom{~~} & Frequency & References \\
 & (erg\,s$^{-1}$\,Hz$^{-1}$) & (GHz) & \\
\hline
  FRB\,20121102A & $1.8$ $\times 10^{29}$ & 1.6 & \citet{Marcote17}\\   
  FRB\,20180916B & $< 4.9 \times 10^{26}$ & 1.7 & \citet{Marcote+20} \\
  FRB\,20181030A & $ < 2.0 \times 10^{26}$ & 3.0 & \citet{Bhardwaj+21} \\
  FRB\,20190520B &$3.0 \times 10^{29}$ & 1.7,\ 3.0 & This work, \citet{Niu+22} \\
  FRB\,20200120E & $< 3.1 \times 10^{23}$ & 1.5 & \citet{Kirsten+22} \\  
  FRB\,20201124A & $< 3.0 \times 10^{28}$ & 1.6 & \citet{RaviR67+22}\\

 \hline
    \end{tabular}
    \caption{ \label{luminosities}PRS spectral radio luminosity measurements and $3-\sigma$ upper limits for a sample of well-localized repeating FRBs at the highest-available angular resolutions.}
\end{table*}

We have measured a PRS source size of $1.4 \pm 0.3\,\mathrm{mas}$ in our EVN data. While this value appears to imply significant source extension, we cannot claim that we are measuring the intrinsic size of the PRS source.
According to \citet{Marti+12}, the smallest resolvable size of a source, $\theta_{\rm min}$, can be expressed as:
\begin{equation}
\theta_{\text{min}} = \beta \left(\frac{\lambda_c}{2}\right)^{1/4} \left( \frac{1}{\textrm{S/N}}\right)^{1/2}\times\theta_{\text{beam}},
\label{OverResEq}
\end{equation}
\noindent where S/N is the signal-to-noise ratio of the averaged visibilities; $\beta$ weakly depends on the spatial distribution of the telescopes (it typically takes values between 0.5 and 1 for VLBI arrays); $\theta_{\text{beam}}$ is the FWHM of the synthesized beam using natural weighting; and $\lambda_c$ depends on the probability cutoff for a false size-detection. The value of $\lambda_c$ is 3.84 for a $2\,\sigma$ cutoff. Equation~\ref{OverResEq} assumes that the source size is estimated directly from the visibilities, by means of model fitting. Following this, we obtain a $\theta_{\rm min} \sim 2$\,mas for $\beta = 1$ for our observations. Our measured size for the PRS is below this limit and thus the source is not actually resolved. We also note that the angular scatter broadening (at 1.7\,GHz) due to the Milky Way ISM along this line of sight is expected to be 0.4\,mas \citep{ne2001} and thus negligible. 

Furthermore, a circular Gaussian fit to the FRB burst emission (assuming the brightest peak being associated with the FRB emission) revealed an apparent source size of $\approx 1.7$--$2\,\mathrm{mas}$. Given that the FRB emission must be compact on much smaller scales, due to the timescale of the burst and causality arguments, we confirm that the measured size of $\sim 2\,\mathrm{mas}$ is not intrinsic but due to the limited resolution of the interferometric data.
Thus, we use the measured size as an upper limit to the intrinsic size of the PRS and constrain its angular size to be $< 2.3\,\mathrm{mas}$ at $3\sigma$ confidence level. Given the redshift of the source ($z=0.241$; \citealt{Niu+22}), this implies a projected physical diameter of $< 9\,\mathrm{pc}$ (at $3\sigma$ confidence level) at an angular diameter distance of 810.5\,Mpc. 
\\
\\

\subsubsection{Comparison with other sources}

Young neutron stars in a supernova remnant (SNR) and/or a pulsar wind nebula (PWN) feature in some of the models proposed for explaining repeating FRBs and PRS emission \citep{Beloborodov+17, Dai+17, Metzger+17, Margalit+18}. 
SNRs are the outcome of prompt energy deposition in the form of a blast wave propelled into the ISM by a supernova explosion. PWNe, on the other hand, have a longer-lived power supply in the form of a bulk relativistic flow of electron/positron pairs from an active neutron star. Moreover, SNRs have relatively steep spectral indices ($-0.8 <\alpha< -0.3$) whereas PWNe have flat spectral indices ($-0.3 <\alpha< 0$). Such a flat spectral index below a frequency of 10\,GHz has been observed for both known FRB PRSs. As far as their physical size is concerned, Galactic SNRs range in size from a few parsecs to a few tens of parsecs, whereas Galactic PWNe are typically parsec-level
\citep{Reynolds+12} --- although some older PWNe may be substantially larger. In addition to SNRs and PWNe, accreting NSs or BHs have also been proposed to explain the high and variable RMs of some repeating FRBs \citep{Sridhar&Metzger22,Sridhar+23b}. Observationally, a sample of compact sources with comparable luminosities as the FRB PRSs have been identified in dwarf galaxies \citep{Reines2020, Vohl2023}. 

Motivated by the size constraints and luminosities of different source types and how they compare to FRB PRSs (see Figure~\ref{fig:LumSize}), we investigate the phase space of radio luminosity and physical size of sources such as the known PRSs associated with FRBs; `wandering' black hole candidates in dwarf galaxies \citep{Reines2020, Sargent2022}; the compact radio source emerging from SN1986J's SNR at the epoch of maximum radio luminosity \citep{SN1986J}; Galactic SNRs and PWNe\footnote{\url{https://www.physics.mcgill.ca/~pulsar/pwncat.html}} \citep{Green2019,Ranasinghe+2023}; SNRs in nearby galaxies with distances ranging from $0.055$--$14.5\,\mathrm{Mpc}$ \citep{eSN}; transient sources such as the PWN candidate VT\,1137$-$0337 \citep{Dong2023} and the orphan GRB afterglow FIRST\,J141918.9$+$394036 \citep{Law18, Marcote2019}; PRS-like candidates identified in a low-frequency survey \citep{Vohl2023}; and low-luminosity AGNe detected with the VLBA spanning a redshift range of $z = 0.3$--$3.4$ \citep{Radcliffe+18}. We note that the sample of Galactic SNRs is nearly complete for remnant ages $<2$\,kyr and shows a mean Galactic SNR diameter of 30.5\,pc \citep{Ranasinghe+2023}.

We scale the radio luminosities of sources to 1\,GHz using their measured spectral indices from the literature. This exclude SNRs and PWNe as their flux densities are measured at 1\,GHz \citep{Ranasinghe+2023}. For sources without a measured spectral index, we assume a canonical value of $\alpha = -0.7$ \citep{Condon+92,Gioia+82}. We find that the radio luminosities of the PRSs are only surpassed by the low-luminosity AGNe, and are about a million times brighter than Galactic SNRs and PWNe, including the Crab nebula. On the other hand, their physical size seems to be broadly consistent with that of other compact radio sources. We note that the physical size of the PRS candidates from \citet{Vohl2023} are poorly constrained due to the limited resolution of the survey (6$^{\prime \prime}$), though ongoing VLBI analysis is being done to better constrain these values. A subset of these and extragalactic SNRs, in addition to the compact radio source in SN1986J, seem to bridge the luminosity gap between the Galactic and extragalactic compact sources. We note that this luminosity gap is most likely a consequence of the present observational biases.

\begin{figure}
\includegraphics[width=\columnwidth]{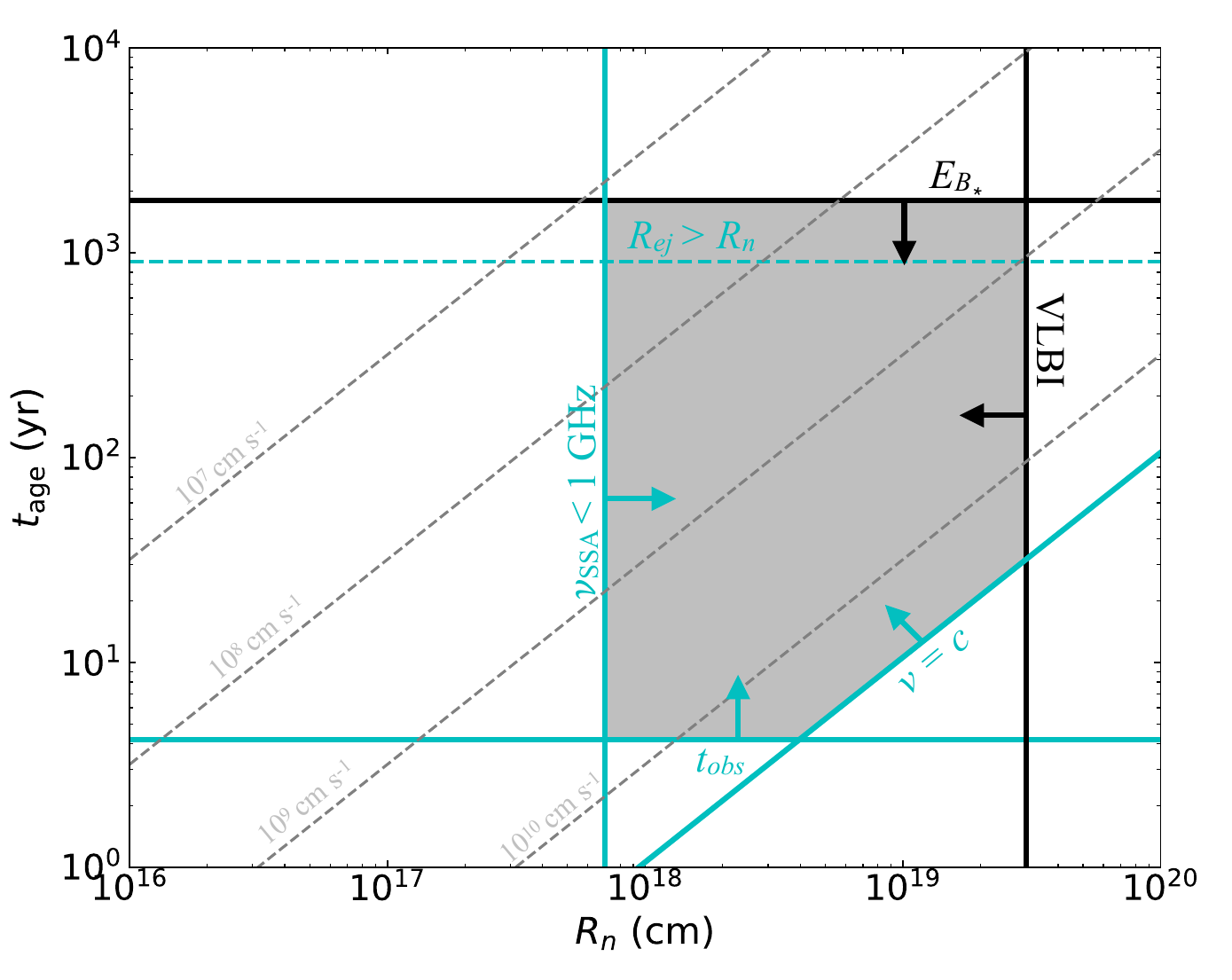}
\caption{Allowed parameter space for the age, $t_{\rm age}$, of a putative magnetar progenitor for FRB\,20190520B and the radius $R_\textit{n}$ of the synchrotron nebula responsible for powering the observed PRS emission, based on observational constraints. An upper limit on $R_\textit{n}$ is obtained from our EVN observations, while a lower limit can be placed based on the lack of a clear self-absorption signature (SSA) below 1\,GHz \citep{Zhang+18}. An upper limit on the source age is obtained from the observed PRS luminosity and total available magnetic energy, while the date of discovery sets an approximate lower limit on the source age. Requiring that the size of the expanding supernova ejecta shell $R_{\textit{ej}}$ is larger than the size of the magnetar-inflated nebula requires a minimum age of $t_{\rm age} \gtrsim 900$\,yr. Dashed gray curves correspond to the expansion velocity of the synchrotron nebula, $R_\textit{n}/t_{\rm age}$.}
\label{fig:magnetar_model}
\end{figure}

\subsection{Progenitor implications}
We examine the model of a magnetar in an ion-wind nebula \citep{Margalit+18} and the hypernebula model \citep{Sridhar&Metzger22,Sridhar+23b} using the luminosity and size constraints for FRB\,20190520B's associated PRS obtained from our EVN observations. 

\subsubsection{Magnetar in a magnetized ion–electron wind nebula}
We place constraints on the age, $t_{\rm age}$, of a putative magnetar responsible for powering both the bursts and PRS associated with FRB\,20190520B following the prescription of \citet{Margalit+18}. In this scenario, the magnetic energy, $E_{\rm B_\thinstar}$, of the magnetar inflates a synchrotron nebula behind the expanding supernova ejecta over a timescale $t_{\rm age}$, 
\begin{equation}
\label{eq:tage}
    t_{\rm age} \lesssim \dfrac{E_{\rm B_\thinstar} }{\nu L_\nu} \approx 1800 \ B^{2}_{16} \ \rm yr
\end{equation}
where $\nu L_\nu \sim 5\times 10^{38} \ \rm erg \ s^{-1}$ is the 1.7-GHz luminosity of the PRS, $B_*$ is the interior magnetic field strength, and $B_{16} = B_*/10^{16} \ \rm G$. Assuming a lower limit on the source age of $t_{\rm age} \gtrsim 4$\,yr based on the initial discovery, we find that the allowed age of $t_{\rm age} \sim 4$--$1900$\,yr is comparable to that inferred for FRB\,20121102A \citep{Margalit+18}, and consistent with the active lifetime of strong $B$-field, millisecond magnetars. Figure~\ref{fig:magnetar_model} shows the allowed parameter space in both $t_{\rm age}$ and $R_n$, the magnetar nebula radius, based on observational constraints, where our EVN data set an upper limit on $R_n$ of $< 9$\,pc, while a lower limit of $\gtrsim 0.22$\,pc can be placed from the lack of a clear break in the synchrotron spectrum due to self-absorption \citep{Zhang+23}. We note, however, that in order to reconcile the EVN limit of $< 9$\,pc for the nebular size with the size of the expanding supernova ejecta shell --- which must be larger than the magnetar-inflated nebula --- requires an age of $\gtrsim 900$\,yr (assuming an ejecta velocity of $v_{\rm ej} \approx 10^4 \ \rm km \ s^{-1}$, typical of hydrogen-poor supernovae). 

\subsubsection{Hypernebula model}
The host's contribution to the overall DM (of $\sim 1177$\,pc\,cm$^{-3}$; see \S\ref{subsec:burst_discovery}) was estimated to be $903^{+72}_{-111}\,{\rm pc\,cm^{-3}}$ \citep{Niu+22}, which is about three times the ${\rm DM}_{\rm host}$ of FRB\,20121102. However, \cite{Lee+23} found that the halos of intervening galaxy groups and clusters contribute significantly to the observed DM, and revised the contribution of the host to ${\rm DM}_{\rm host}=430^{+140}_{-220}\,{\rm pc\,cm^{-3}}$. Based on the FRB scattering timescale, the gas contributing to this ${\rm DM}_{\rm host}$ is also expected to be close to the FRB engine. Such large ${\rm DM}_{\rm host}$ and RM values for FRB\,20190520B ($\sim2\times10^5\,{\rm rad\,m^{-2}}$; \citealt{Niu+22}) hint toward a dense, young, and highly magnetized circum-source medium: conditions naturally expected from a young hypernebula inflated by an evolved accreting binary system. In this scenario, the peak burst luminosity from FRB\,20190520B of $\sim10^{42}\,{\rm erg\,s^{-1}}$ would require a central engine accreting at least at a rate $\overset{\boldsymbol{\cdot}}{\rm M}\gtrsim10^5\overset{\boldsymbol{\cdot}}{\rm M}_{\rm Edd,10}$, where $\overset{\boldsymbol{\cdot}}{\rm M}_{\rm Edd,10}$ is the Eddington-limited mass transfer rate for a black hole with a mass of $10\,{\rm M}_\odot$ \citep{Sridhar+21b}. Locally, the shell inflated by the mass loss in the form of disk winds (at a rate comparable to $\overset{\boldsymbol{\cdot}}{\rm M}$ with a speed $v_{\rm w}\sim0.03\,c$) could contribute as much as \citep{Sridhar&Metzger22},
\begin{equation}
{\rm DM}_{\rm local} \sim{425}\,{\rm pc\,cm^{-3}} \left(\frac{\overset{\boldsymbol{\cdot}}{\rm M}}{{5\times10^5}\overset{\boldsymbol{\cdot}}{\rm M}_{\rm Edd,10}}\right)\left(\frac{v_{\rm w}}{0.03\,c}\right)^{-2}\left(\frac{t_{\rm age}}{10\,{\rm yr}}\right)^{-1}
\end{equation}
to the ${\rm DM}_{\rm host}$ after $t_{\rm age}=10$\,yr of free expansion (i.e., prior to the shell decelerating). This is consistent with the observed ${\rm DM}_{\rm host}$, assuming that the interstellar medium of the host galaxy contributes $\sim100\,{\rm pc\,cm^{-3}}$ to the ${\rm DM}_{\rm host}$. At this age, the freely-expanding shell would have expanded to a size,
\begin{equation}\label{eq:Rs}
R_{\rm s} \sim 0.1\,{\rm pc} \left(\frac{v_{\rm w}}{0.03\,c}\right)\left(\frac{t_{\rm age}}{10\,{\rm yr}}\right),
\end{equation}
with a central shock-heated nebular core of size $\sim0.01$\,pc \citep[Equation~28 of][]{Sridhar&Metzger22}. We note that these size estimates (Equation~\ref{eq:Rs}) are in agreement with the observed upper limit on the size of the PRS being $<9$\,pc, as reported in this work (\S\ref{subsec:PRS}).  

Recently, \cite{Zhang+23} detected not only a likely radio flux density decrease, but also a marginal variability in the flux density of the PRS associated with FRB\,20190520B between the observations taken in 2020 and 2021 at 3\,GHz. If the variability is attributed to scintillation by a scattering disk (estimated to be of angular size $53\,\upmu{\rm as}$ at 3\,GHz based on the observations separated by more than one year) located at a distance of 1\,kpc away from the observer, the size of the PRS is $\sim0.2$\,pc; this drops down to $\sim0.07$\,pc for a scattering disk at 10\,kpc. These estimates are comparable to the size obtained from our modeling (see Equation~\ref{eq:Rs}), and the inference of \cite{Zhang+23} that accreting compact objects such as hypernebulae might be able to explain the PRS’s temporal and spectral properties aligns with ours. Furthermore, as shown in \S4.2 of \cite{Sridhar&Metzger22}, this model also self-consistently explains the observed luminosity of the PRS ($L_{1.7\,\rm GHz} \sim 3 \times 10^{29}\,\mathrm{erg\,s^{-1}\,Hz^{-1}}$), and the large and varying RM of the FRBs ($\sim2\times10^5\,{\rm rad\,m^{-2}}$; \citealt{Niu+22}) as they traverse through the hypernebula.

\section{Conclusions} \label{section:conclusions}
We have presented tight constraints on the position and size of the radio source associated with FRB\,20190520B and confirmed it to be an FRB PRS by characterizing its compact nature using EVN observations. We conclude the following:
\begin{enumerate}
    \item We have detected a PRS with an angular size of $<~2.3$\,mas ($3\sigma$). This results in a physical diameter of $<9$\,pc at a angular diameter distance of 810.5\,Mpc.
    
    \item We find the position of the PRS to be: \\ $\alpha (\text{J}2000) = 16^{\rm h}02^{\rm m}04.2611^{\rm s} \pm 6.5\,\rm{mas}$,\\ 
    $\delta (\text{J}2000) = -11^\circ17^\prime17.366^{\prime\prime} \pm 3.6~\mathrm{mas}$\\
    referenced to the ICRF.
    This position is consistent within uncertainties with the previously published VLA position of \citet{Niu+22} and is $15\times$ more precise.
    
    \item We have detected and localized a burst from  FRB\,20190520B during our EVN-PRECISE observations. Though we cannot independently measure the burst position with the same accuracy as the PRS, we find that the two are consistent with being co-located to within $\lesssim20$\,mas (i.e., to within a transverse distance of $\lesssim80$\,pc).
    
    \item We find the flux density of the PRS to be $201 \pm 34\,\mathrm{\upmu Jy}$ at 1.7\,GHz. The flux density is consistent between two epochs separated by a day and also consistent with other published values \citep{Niu+22,Zhang+23}. As a result, we can confirm that no flux density is resolved at VLBI scales. Using the measured flux density, we find the spectral radio luminosity of the PRS to be $L_{1.7\,\rm GHz} = (3.0 \pm 0.5) \times 10^{29}\,\mathrm{erg\,s^{-1}\,Hz^{-1}}$ at the given luminosity distance of 1.25\,Gpc. Our luminosity is also consistent with \citet{Niu+22}.

    \item Based on our EVN observations and results, we have explored the model of a magnetar in a magnetized ion–electron wind nebula \citep{Margalit+18}. We have presented an allowed parameter space for the age of a putative magnetar progenitor ($4-1900$\,yr) for FRB\,20190520B and the radius of the synchroton nebula responsible for powering the observed PRS emission. Furthermore, we also considered the accretion-powered `hypernebula' model \citep{Sridhar&Metzger22, Sridhar+23b} in light of our results and find that the model estimates for the nebular size, PRS luminosity, rotation measure, and the host's contribution to the DM to be in agreement with our observational constraints.
\end{enumerate}
Further observations are currently being conducted within our EVN-PRECISE program. The detection of a larger number of bursts will allow us to pin-point the burst source to milliarcsecond precision, independently of any information on the position of the PRS. This will more robustly constrain its physical separation from the PRS at the parsec level.
\\
\\

We would like to thank the directors and staff at the various participating stations for allowing us to use their facilities and running the observations. SB would like to thank Dany Vohl for providing a list of a PRS-like candidate and Pragya Chawla for useful discussions. 
The European VLBI Network is a joint facility of independent European, African, Asian, and North American radio astronomy institutes. Scientific results from data presented in this publication are derived from the following EVN project codes: EK051, EM161.
We note that the antennas T6 and Ur originally observed linear polarizations in the EM161 observations, which were transformed to a circular basis during the internal EVN post-processing using the PolConvert program \citep{polcol}.
The research leading to these results has received funding from the European Union's Horizon 2020 Research and Innovation Programme under grant agreement No. 101004719 (OPTICON RadioNet Pilot).

SB is supported by a Dutch Research Council (NWO) Veni Fellowship (VI.Veni.212.058).
Research by the AstroFlash group at University of Amsterdam, ASTRON and
JIVE is supported in part by an NWO Vici grant (PI Hessels; VI.C.192.045).
BM acknowledges financial support from the State Agency for Research of the Spanish Ministry of Science and Innovation under grant PID2019-105510GB-C31/AEI/10.13039/501100011033 and through the Unit of Excellence Mar\'ia de Maeztu 2020--2023 award to the Institute of Cosmos Sciences (CEX2019-000918-M).
FK acknowledges support from Onsala Space Observatory for the provisioning of its facilities/observational support. The Onsala Space Observatory national research infrastructure is funded through Swedish Research Council grant No 2017-00648.
NS acknowledges the support from NASA (grant number 80NSSC22K0332), NASA FINESST (grant number 80NSSC22K1597), Columbia University Dean’s fellowship, and a grant from the Simons Foundation.
NS performed part of this work at the Aspen Center for Physics, which is supported by the National Science Foundation grant PHY-2210452.
TE is supported by NASA through the NASA Hubble Fellowship grant HST-HF2-51504.001-A awarded by the Space Telescope Science Institute, which is operated by the Association of Universities for Research in Astronomy, Inc., for NASA, under contract NAS5-26555.
\facilities{EVN}
\software{SFXC \citep{Keimpema+15}, APLpy \citep{aplpy}, Astropy \citep{astropy1}, CASA \citep{CASA}, AIPS \citep{AIPS}, Difmap \citep{DIFMAP}, DSPSR \citep{vanstraten_2011_pasa}, PSRCHIVE \citep{vanstraten_2012_art, hotan_2004_pasa}, Sigproc \citep{lorimer_2011_ascl}. }  
\bibliography{references_new}{}

\begin{thebibliography}{}
\expandafter\ifx\csname natexlab\endcsname\relax\def\natexlab#1{#1}\fi
\providecommand{\url}[1]{\href{#1}{#1}}
\providecommand{\dodoi}[1]{doi:~\href{http://doi.org/#1}{\nolinkurl{#1}}}
\providecommand{\doeprint}[1]{\href{http://ascl.net/#1}{\nolinkurl{http://ascl.net/#1}}}
\providecommand{\doarXiv}[1]{\href{https://arxiv.org/abs/#1}{\nolinkurl{https://arxiv.org/abs/#1}}}

\bibitem[{{Agarwal} {et~al.}(2020){Agarwal}, {Aggarwal}, {Burke-Spolaor},
  {Lorimer}, \& {Garver-Daniels}}]{Agarwal:2020}
{Agarwal}, D., {Aggarwal}, K., {Burke-Spolaor}, S., {Lorimer}, D.~R., \&
  {Garver-Daniels}, N. 2020, \mnras, 497, 1661, \dodoi{10.1093/mnras/staa1856}

\bibitem[{{Amiri} {et~al.}(2021){Amiri}, {Andersen}, {Bandura}, {Berger},
  {Bhardwaj}, {Boyce}, {Boyle}, {Brar}, {Breitman}, {Cassanelli}, {Chawla},
  {Chen}, {Cliche}, {Cook}, {Cubranic}, {Curtin}, {Deng}, {Dobbs}, {(Adam)
  Dong}, {Eadie}, {Fandino}, {Fonseca}, {Gaensler}, {Giri}, {Good}, {Halpern},
  {Hill}, {Hinshaw}, {Josephy}, {Kaczmarek}, {Kader}, {Kania}, {Kaspi},
  {Landecker}, {Lang}, {Leung}, {Li}, {Lin}, {Masui}, {McKinven}, {Mena-Parra},
  {Merryfield}, {Meyers}, {Michilli}, {Milutinovic}, {Mirhosseini},
  {M{\"u}nchmeyer}, {Naidu}, {Newburgh}, {Ng}, {Patel}, {Pen}, {Petroff},
  {Pinsonneault-Marotte}, {Pleunis}, {Rafiei-Ravandi}, {Rahman}, {Ransom},
  {Renard}, {Sanghavi}, {Scholz}, {Shaw}, {Shin}, {Siegel}, {Sikora}, {Singh},
  {Smith}, {Stairs}, {Tan}, {Tendulkar}, {Vanderlinde}, {Wang}, {Wulf},
  {Zwaniga}, \& {CHIME/FRB Collaboration}}]{chimecat}
{Amiri}, M., {Andersen}, B.~C., {Bandura}, K., {et~al.} 2021, \apjs, 257, 59,
  \dodoi{10.3847/1538-4365/ac33ab}

\bibitem[{{Anna-Thomas} {et~al.}(2022){Anna-Thomas}, {Connor}, {Burke-Spolaor},
  {Beniamini}, {Aggarwal}, {Law}, {Lynch}, {Li}, {Feng}, {Ocker}, {Cruces},
  {Chatterjee}, {Yu}, {Niu}, \& {Xue}}]{Anna-Thomas+22}
{Anna-Thomas}, R., {Connor}, L., {Burke-Spolaor}, S., {et~al.} 2022, arXiv
  e-prints, arXiv:2202.11112.
\newblock \doarXiv{2202.11112}

\bibitem[{{Beloborodov}(2017)}]{Beloborodov+17}
{Beloborodov}, A.~M. 2017, \apjl, 843, L26, \dodoi{10.3847/2041-8213/aa78f3}

\bibitem[{{Bhandari} {et~al.}(2020){Bhandari}, {Bannister}, {Lenc}, {Cho},
  {Ekers}, {Day}, {Deller}, {Flynn}, {James}, {Macquart}, {Mahony}, {Marnoch},
  {Moss}, {Phillips}, {Prochaska}, {Qiu}, {Ryder}, {Shannon}, {Tejos}, \&
  {Wong}}]{Bhandari+20}
{Bhandari}, S., {Bannister}, K.~W., {Lenc}, E., {et~al.} 2020, \apjl, 901, L20,
  \dodoi{10.3847/2041-8213/abb462}

\bibitem[{{Bhandari} {et~al.}(2022){Bhandari}, {Heintz}, {Aggarwal}, {Marnoch},
  {Day}, {Sydnor}, {Burke-Spolaor}, {Law}, {Xavier Prochaska}, {Tejos},
  {Bannister}, {Butler}, {Deller}, {Ekers}, {Flynn}, {Fong}, {James}, {Lazio},
  {Luo}, {Mahony}, {Ryder}, {Sadler}, {Shannon}, {Han}, {Lee}, \&
  {Zhang}}]{Bhandari+22}
{Bhandari}, S., {Heintz}, K.~E., {Aggarwal}, K., {et~al.} 2022, \aj, 163, 69,
  \dodoi{10.3847/1538-3881/ac3aec}

\bibitem[{{Bhardwaj} {et~al.}(2021){Bhardwaj}, {Kirichenko}, {Michilli},
  {Mayya}, {Kaspi}, {Gaensler}, {Rahman}, {Tendulkar}, {Fonseca}, {Josephy},
  {Leung}, {Merryfield}, {Petroff}, {Pleunis}, {Sanghavi}, {Scholz}, {Shin},
  {Smith}, \& {Stairs}}]{Bhardwaj+21}
{Bhardwaj}, M., {Kirichenko}, A.~Y., {Michilli}, D., {et~al.} 2021, \apjl, 919,
  L24, \dodoi{10.3847/2041-8213/ac223b}

\bibitem[{{Bietenholz} \& {Bartel}(2017)}]{SN1986J}
{Bietenholz}, M.~F., \& {Bartel}, N. 2017, \apj, 839, 10,
  \dodoi{10.3847/1538-4357/aa67a0}

\bibitem[{{Bochenek} {et~al.}(2020){Bochenek}, {Ravi}, {Belov}, {Hallinan},
  {Kocz}, {Kulkarni}, \& {McKenna}}]{Bochenek+20}
{Bochenek}, C.~D., {Ravi}, V., {Belov}, K.~V., {et~al.} 2020, \nat, 587, 59,
  \dodoi{10.1038/s41586-020-2872-x}

\bibitem[{{Chatterjee} {et~al.}(2017){Chatterjee}, {Law}, {Wharton},
  {Burke-Spolaor}, {Hessels}, {Bower}, {Cordes}, {Tendulkar}, {Bassa},
  {Demorest}, {Butler}, {Seymour}, {Scholz}, {Abruzzo}, {Bogdanov}, {Kaspi},
  {Keimpema}, {Lazio}, {Marcote}, {McLaughlin}, {Paragi}, {Ransom}, {Rupen},
  {Spitler}, \& {van Langevelde}}]{VLAlocalisation}
{Chatterjee}, S., {Law}, C.~J., {Wharton}, R.~S., {et~al.} 2017, \nat, 541, 58,
  \dodoi{10.1038/nature20797}

\bibitem[{{Chen} {et~al.}(2022){Chen}, {Ravi}, \& {Hallinan}}]{Chen+22}
{Chen}, G., {Ravi}, V., \& {Hallinan}, G.~W. 2022, arXiv e-prints,
  arXiv:2201.00999, \dodoi{10.48550/arXiv.2201.00999}

\bibitem[{{CHIME/FRB Collaboration} {et~al.}(2020){CHIME/FRB Collaboration},
  {Andersen}, {Bandura}, {Bhardwaj}, {Bij}, {Boyce}, {Boyle}, {Brar},
  {Cassanelli}, {Chawla}, {Chen}, {Cliche}, {Cook}, {Cubranic}, {Curtin},
  {Denman}, {Dobbs}, {Dong}, {Fandino}, {Fonseca}, {Gaensler}, {Giri}, {Good},
  {Halpern}, {Hill}, {Hinshaw}, {H{\"o}fer}, {Josephy}, {Kania}, {Kaspi},
  {Landecker}, {Leung}, {Li}, {Lin}, {Masui}, {McKinven}, {Mena-Parra},
  {Merryfield}, {Meyers}, {Michilli}, {Milutinovic}, {Mirhosseini},
  {M{\"u}nchmeyer}, {Naidu}, {Newburgh}, {Ng}, {Patel}, {Pen},
  {Pinsonneault-Marotte}, {Pleunis}, {Quine}, {Rafiei-Ravandi}, {Rahman},
  {Ransom}, {Renard}, {Sanghavi}, {Scholz}, {Shaw}, {Shin}, {Siegel}, {Singh},
  {Smegal}, {Smith}, {Stairs}, {Tan}, {Tendulkar}, {Tretyakov}, {Vanderlinde},
  {Wang}, {Wulf}, \& {Zwaniga}}]{2020Natur.587...54C}
{CHIME/FRB Collaboration}, {Andersen}, B.~C., {Bandura}, K.~M., {et~al.} 2020,
  \nat, 587, 54, \dodoi{10.1038/s41586-020-2863-y}

\bibitem[{{Condon}(1992)}]{Condon+92}
{Condon}, J.~J. 1992, \araa, 30, 575,
  \dodoi{10.1146/annurev.aa.30.090192.003043}

\bibitem[{{Cordes} \& {Lazio}(2002)}]{ne2001}
{Cordes}, J.~M., \& {Lazio}, T.~J.~W. 2002, arXiv e-prints, arXiv:0207156.
\newblock \doarXiv{astro-ph/0207156}

\bibitem[{{Dai} {et~al.}(2017){Dai}, {Wang}, \& {Yu}}]{Dai+17}
{Dai}, Z.~G., {Wang}, J.~S., \& {Yu}, Y.~W. 2017, \apjl, 838, L7,
  \dodoi{10.3847/2041-8213/aa6745}

\bibitem[{{Dong} \& {Hallinan}(2023)}]{Dong2023}
{Dong}, D.~Z., \& {Hallinan}, G. 2023, \apj, 948, 119,
  \dodoi{10.3847/1538-4357/acc06c}

\bibitem[{{Dong} {et~al.}(2023){Dong}, {Eftekhari}, {Fong}, {Deller},
  {Mannings}, {Simha}, {Sridhar}, {Rafelski}, {Gordon}, {Bhandari}, {Day},
  {Heintz}, {Hessels}, {Leja}, {James}, {Kilpatrick}, {Mahony}, {Marcote},
  {Margalit}, {Nimmo}, {Prochaska}, {Rouco Escorial}, {Ryder}, {Schroeder},
  {Shannon}, \& {Tejos}}]{Dong+23}
{Dong}, Y., {Eftekhari}, T., {Fong}, W.-f., {et~al.} 2023, arXiv e-prints,
  arXiv:2307.06995, \dodoi{10.48550/arXiv.2307.06995}

\bibitem[{{Gioia} {et~al.}(1982){Gioia}, {Gregorini}, \& {Klein}}]{Gioia+82}
{Gioia}, I.~M., {Gregorini}, L., \& {Klein}, U. 1982, \aap, 116, 164

\bibitem[{{Gordon} {et~al.}(2023){Gordon}, {Fong}, {Kilpatrick}, {Eftekhari},
  {Leja}, {Prochaska}, {Nugent}, {Bhandari}, {Blanchard}, {Caleb}, {Day},
  {Deller}, {Dong}, {Glowacki}, {Gourdji}, {Mannings}, {Mahoney}, {Marnoch},
  {Miller}, {Paterson}, {Rastinejad}, {Ryder}, {Sadler}, {Scott}, {Sears},
  {Shannon}, {Simha}, {Stappers}, \& {Tejos}}]{Gordon+23}
{Gordon}, A.~C., {Fong}, W.-f., {Kilpatrick}, C.~D., {et~al.} 2023, arXiv
  e-prints, arXiv:2302.05465, \dodoi{10.48550/arXiv.2302.05465}

\bibitem[{{Gourdji} {et~al.}(2019){Gourdji}, {Michilli}, {Spitler}, {Hessels},
  {Seymour}, {Cordes}, \& {Chatterjee}}]{Gourdji+19}
{Gourdji}, K., {Michilli}, D., {Spitler}, L.~G., {et~al.} 2019, \apjl, 877,
  L19, \dodoi{10.3847/2041-8213/ab1f8a}

\bibitem[{{Green}(2019)}]{Green2019}
{Green}, D.~A. 2019, Journal of Astrophysics and Astronomy, 40, 36,
  \dodoi{10.1007/s12036-019-9601-6}

\bibitem[{{Greisen}(2003)}]{AIPS}
{Greisen}, E.~W. 2003, in Astrophysics and Space Science Library, Vol. 285,
  Information Handling in Astronomy - Historical Vistas, ed. A.~{Heck}, 109,
  \dodoi{10.1007/0-306-48080-8_7}

\bibitem[{{Hilmarsson} {et~al.}(2021){Hilmarsson}, {Michilli}, {Spitler},
  {Wharton}, {Demorest}, {Desvignes}, {Gourdji}, {Hackstein}, {Hessels},
  {Nimmo}, {Seymour}, {Kramer}, \& {Mckinven}}]{Hilmarsson+21}
{Hilmarsson}, G.~H., {Michilli}, D., {Spitler}, L.~G., {et~al.} 2021, \apjl,
  908, L10, \dodoi{10.3847/2041-8213/abdec0}

\bibitem[{{Hotan} {et~al.}(2004){Hotan}, {van Straten}, \&
  {Manchester}}]{hotan_2004_pasa}
{Hotan}, A.~W., {van Straten}, W., \& {Manchester}, R.~N. 2004, \pasa, 21, 302,
  \dodoi{10.1071/AS04022}

\bibitem[{{James} {et~al.}(2022){James}, {Ghosh}, {Prochaska}, {Bannister},
  {Bhandari}, {Day}, {Deller}, {Glowacki}, {Gordon}, {Heintz}, {Marnoch},
  {Ryder}, {Scott}, {Shannon}, \& {Tejos}}]{James+22b}
{James}, C.~W., {Ghosh}, E.~M., {Prochaska}, J.~X., {et~al.} 2022, \mnras, 516,
  4862, \dodoi{10.1093/mnras/stac2524}

\bibitem[{{Keimpema} {et~al.}(2015){Keimpema}, {Kettenis}, {Pogrebenko},
  {Campbell}, {Cim{\'o}}, {Duev}, {Eldering}, {Kruithof}, {van Langevelde},
  {Marchal}, {Molera Calv{\'e}s}, {Ozdemir}, {Paragi}, {Pidopryhora},
  {Szomoru}, \& {Yang}}]{Keimpema+15}
{Keimpema}, A., {Kettenis}, M.~M., {Pogrebenko}, S.~V., {et~al.} 2015,
  Experimental Astronomy, 39, 259, \dodoi{10.1007/s10686-015-9446-1}

\bibitem[{{Kirsten} {et~al.}(2021){Kirsten}, {Snelders}, {Jenkins}, {Nimmo},
  {van den Eijnden}, {Hessels}, {Gawro{\'n}ski}, \& {Yang}}]{Kirsten+21}
{Kirsten}, F., {Snelders}, M.~P., {Jenkins}, M., {et~al.} 2021, Nature
  Astronomy, 5, 414, \dodoi{10.1038/s41550-020-01246-3}

\bibitem[{{Kirsten} {et~al.}(2015){Kirsten}, {Vlemmings}, {Campbell}, {Kramer},
  \& {Chatterjee}}]{Kirsten+15}
{Kirsten}, F., {Vlemmings}, W., {Campbell}, R.~M., {Kramer}, M., \&
  {Chatterjee}, S. 2015, \aap, 577, A111, \dodoi{10.1051/0004-6361/201425562}

\bibitem[{{Kirsten} {et~al.}(2022){Kirsten}, {Marcote}, {Nimmo}, {Hessels},
  {Bhardwaj}, {Tendulkar}, {Keimpema}, {Yang}, {Snelders}, {Scholz},
  {Pearlman}, {Law}, {Peters}, {Giroletti}, {Paragi}, {Bassa}, {Hewitt},
  {Bach}, {Bezrukovs}, {Burgay}, {Buttaccio}, {Conway}, {Corongiu}, {Feiler},
  {Forss{\'e}n}, {Gawro{\'n}ski}, {Karuppusamy}, {Kharinov}, {Lindqvist},
  {Maccaferri}, {Melnikov}, {Ould-Boukattine}, {Possenti}, {Surcis}, {Wang},
  {Yuan}, {Aggarwal}, {Anna-Thomas}, {Bower}, {Blaauw}, {Burke-Spolaor},
  {Cassanelli}, {Clarke}, {Fonseca}, {Gaensler}, {Gopinath}, {Kaspi}, {Kassim},
  {Lazio}, {Leung}, {Li}, {Lin}, {Masui}, {Mckinven}, {Michilli}, {Mikhailov},
  {Ng}, {Orbidans}, {Pen}, {Petroff}, {Rahman}, {Ransom}, {Shin}, {Smith},
  {Stairs}, \& {Vlemmings}}]{Kirsten+22}
{Kirsten}, F., {Marcote}, B., {Nimmo}, K., {et~al.} 2022, \nat, 602, 585,
  \dodoi{10.1038/s41586-021-04354-w}

\bibitem[{{Law} {et~al.}(2022){Law}, {Connor}, \& {Aggarwal}}]{Law+22}
{Law}, C.~J., {Connor}, L., \& {Aggarwal}, K. 2022, \apj, 927, 55,
  \dodoi{10.3847/1538-4357/ac4c42}

\bibitem[{{Law} {et~al.}(2018){Law}, {Bower}, {Burke-Spolaor}, {Butler},
  {Demorest}, {Halle}, {Khudikyan}, {Lazio}, {Pokorny}, {Robnett}, \&
  {Rupen}}]{Law18}
{Law}, C.~J., {Bower}, G.~C., {Burke-Spolaor}, S., {et~al.} 2018, \apjs, 236,
  8, \dodoi{10.3847/1538-4365/aab77b}

\bibitem[{{Lee} {et~al.}(2023){Lee}, {Khrykin}, {Simha}, {Ata}, {Huang},
  {Prochaska}, {Tejos}, {Cooke}, {Nagamine}, \& {Zhang}}]{Lee+23}
{Lee}, K.-G., {Khrykin}, I.~S., {Simha}, S., {et~al.} 2023, \apjl, 954, L7,
  \dodoi{10.3847/2041-8213/acefb5}

\bibitem[{{Lorimer}(2011{\natexlab{a}})}]{Sigproc}
{Lorimer}, D.~R. 2011{\natexlab{a}}, {SIGPROC: Pulsar Signal Processing
  Programs}, Astrophysics Source Code Library, record ascl:1107.016.
\newblock \doeprint{1107.016}

\bibitem[{{Lorimer}(2011{\natexlab{b}})}]{lorimer_2011_ascl}
---. 2011{\natexlab{b}}, {SIGPROC: Pulsar Signal Processing Programs},
  Astrophysics Source Code Library, record ascl:1107.016.
\newblock \doeprint{1107.016}

\bibitem[{{Marcote} {et~al.}(2019){Marcote}, {Nimmo}, {Salafia}, {Paragi},
  {Hessels}, {Petroff}, \& {Karuppusamy}}]{Marcote2019}
{Marcote}, B., {Nimmo}, K., {Salafia}, O.~S., {et~al.} 2019, \apjl, 876, L14,
  \dodoi{10.3847/2041-8213/ab1aad}

\bibitem[{{Marcote} {et~al.}(2017){Marcote}, {Paragi}, {Hessels}, {Keimpema},
  {van Langevelde}, {Huang}, {Bassa}, {Bogdanov}, {Bower}, {Burke-Spolaor},
  {Butler}, {Campbell}, {Chatterjee}, {Cordes}, {Demorest}, {Garrett}, {Ghosh},
  {Kaspi}, {Law}, {Lazio}, {McLaughlin}, {Ransom}, {Salter}, {Scholz},
  {Seymour}, {Siemion}, {Spitler}, {Tendulkar}, \& {Wharton}}]{Marcote17}
{Marcote}, B., {Paragi}, Z., {Hessels}, J.~W.~T., {et~al.} 2017, \apjl, 834,
  L8, \dodoi{10.3847/2041-8213/834/2/L8}

\bibitem[{{Marcote} {et~al.}(2020){Marcote}, {Nimmo}, {Hessels}, {Tendulkar},
  {Bassa}, {Paragi}, {Keimpema}, {Bhardwaj}, {Karuppusamy}, {Kaspi}, {Law},
  {Michilli}, {Aggarwal}, {Andersen}, {Archibald}, {Bandura}, {Bower}, {Boyle},
  {Brar}, {Burke-Spolaor}, {Butler}, {Cassanelli}, {Chawla}, {Demorest},
  {Dobbs}, {Fonseca}, {Giri}, {Good}, {Gourdji}, {Josephy}, {Kirichenko},
  {Kirsten}, {Landecker}, {Lang}, {Lazio}, {Li}, {Lin}, {Linford}, {Masui},
  {Mena-Parra}, {Naidu}, {Ng}, {Patel}, {Pen}, {Pleunis}, {Rafiei-Ravandi},
  {Rahman}, {Renard}, {Scholz}, {Siegel}, {Smith}, {Stairs}, {Vanderlinde}, \&
  {Zwaniga}}]{Marcote+20}
{Marcote}, B., {Nimmo}, K., {Hessels}, J.~W.~T., {et~al.} 2020, \nat, 577, 190,
  \dodoi{10.1038/s41586-019-1866-z}

\bibitem[{{Margalit} \& {Metzger}(2018)}]{Margalit+18}
{Margalit}, B., \& {Metzger}, B.~D. 2018, \apjl, 868, L4,
  \dodoi{10.3847/2041-8213/aaedad}

\bibitem[{{Mart{\'\i}-Vidal} {et~al.}(2012){Mart{\'\i}-Vidal},
  {P{\'e}rez-Torres}, \& {Lobanov}}]{Marti+12}
{Mart{\'\i}-Vidal}, I., {P{\'e}rez-Torres}, M.~A., \& {Lobanov}, A.~P. 2012,
  \aap, 541, A135, \dodoi{10.1051/0004-6361/201118334}

\bibitem[{{Mart{\'\i}-Vidal} {et~al.}(2016){Mart{\'\i}-Vidal}, {Roy}, {Conway},
  \& {Zensus}}]{polcol}
{Mart{\'\i}-Vidal}, I., {Roy}, A., {Conway}, J., \& {Zensus}, A.~J. 2016, \aap,
  587, A143, \dodoi{10.1051/0004-6361/201526063}

\bibitem[{{McMullin} {et~al.}(2007){McMullin}, {Waters}, {Schiebel}, {Young},
  \& {Golap}}]{CASA}
{McMullin}, J.~P., {Waters}, B., {Schiebel}, D., {Young}, W., \& {Golap}, K.
  2007, in Astronomical Society of the Pacific Conference Series, Vol. 376,
  Astronomical Data Analysis Software and Systems XVI, ed. R.~A. {Shaw},
  F.~{Hill}, \& D.~J. {Bell}, 127

\bibitem[{{Metzger} {et~al.}(2017){Metzger}, {Berger}, \&
  {Margalit}}]{Metzger+17}
{Metzger}, B.~D., {Berger}, E., \& {Margalit}, B. 2017, \apj, 841, 14,
  \dodoi{10.3847/1538-4357/aa633d}

\bibitem[{{Michilli} {et~al.}(2018){Michilli}, {Seymour}, {Hessels}, {Spitler},
  {Gajjar}, {Archibald}, {Bower}, {Chatterjee}, {Cordes}, {Gourdji}, {Heald},
  {Kaspi}, {Law}, {Sobey}, {Adams}, {Bassa}, {Bogdanov}, {Brinkman},
  {Demorest}, {Fernandez}, {Hellbourg}, {Lazio}, {Lynch}, {Maddox}, {Marcote},
  {McLaughlin}, {Paragi}, {Ransom}, {Scholz}, {Siemion}, {Tendulkar}, {van
  Rooy}, {Wharton}, \& {Whitlow}}]{MichilliRM}
{Michilli}, D., {Seymour}, A., {Hessels}, J.~W.~T., {et~al.} 2018, \nat, 553,
  182, \dodoi{10.1038/nature25149}

\bibitem[{{Nimmo} {et~al.}(2022){Nimmo}, {Hewitt}, {Hessels}, {Kirsten},
  {Marcote}, {Bach}, {Blaauw}, {Burgay}, {Corongiu}, {Feiler}, {Gawro{\'n}ski},
  {Giroletti}, {Karuppusamy}, {Keimpema}, {Kharinov}, {Lindqvist},
  {Maccaferri}, {Melnikov}, {Mikhailov}, {Ould-Boukattine}, {Paragi}, {Pilia},
  {Possenti}, {Snelders}, {Surcis}, {Trudu}, {Venturi}, {Vlemmings}, {Wang},
  {Yang}, \& {Yuan}}]{Nimmo+22b}
{Nimmo}, K., {Hewitt}, D.~M., {Hessels}, J.~W.~T., {et~al.} 2022, \apjl, 927,
  L3, \dodoi{10.3847/2041-8213/ac540f}

\bibitem[{{Niu} {et~al.}(2022){Niu}, {Aggarwal}, {Li}, {Zhang}, {Chatterjee},
  {Tsai}, {Yu}, {Law}, {Burke-Spolaor}, {Cordes}, {Zhang}, {Ocker}, {Yao},
  {Wang}, {Feng}, {Niino}, {Bochenek}, {Cruces}, {Connor}, {Jiang}, {Dai},
  {Luo}, {Li}, {Miao}, {Niu}, {Anna-Thomas}, {Sydnor}, {Stern}, {Wang}, {Yuan},
  {Yue}, {Zhou}, {Yan}, {Zhu}, \& {Zhang}}]{Niu+22}
{Niu}, C.~H., {Aggarwal}, K., {Li}, D., {et~al.} 2022, \nat, 606, 873,
  \dodoi{10.1038/s41586-022-04755-5}

\bibitem[{{Petroff} {et~al.}(2022){Petroff}, {Hessels}, \&
  {Lorimer}}]{Petroff+22}
{Petroff}, E., {Hessels}, J.~W.~T., \& {Lorimer}, D.~R. 2022, \aapr, 30, 2,
  \dodoi{10.1007/s00159-022-00139-w}

\bibitem[{{Planck Collaboration} {et~al.}(2020){Planck Collaboration},
  {Aghanim}, {Akrami}, {Ashdown}, {Aumont}, {Baccigalupi}, {Ballardini},
  {Banday}, {Barreiro}, {Bartolo}, {Basak}, {Battye}, {Benabed}, {Bernard},
  {Bersanelli}, {Bielewicz}, {Bock}, {Bond}, {Borrill}, {Bouchet}, {Boulanger},
  {Bucher}, {Burigana}, {Butler}, {Calabrese}, {Cardoso}, {Carron},
  {Challinor}, {Chiang}, {Chluba}, {Colombo}, {Combet}, {Contreras}, {Crill},
  {Cuttaia}, {de Bernardis}, {de Zotti}, {Delabrouille}, {Delouis}, {Di
  Valentino}, {Diego}, {Dor{\'e}}, {Douspis}, {Ducout}, {Dupac}, {Dusini},
  {Efstathiou}, {Elsner}, {En{\ss}lin}, {Eriksen}, {Fantaye}, {Farhang},
  {Fergusson}, {Fernandez-Cobos}, {Finelli}, {Forastieri}, {Frailis},
  {Fraisse}, {Franceschi}, {Frolov}, {Galeotta}, {Galli}, {Ganga},
  {G{\'e}nova-Santos}, {Gerbino}, {Ghosh}, {Gonz{\'a}lez-Nuevo}, {G{\'o}rski},
  {Gratton}, {Gruppuso}, {Gudmundsson}, {Hamann}, {Handley}, {Hansen},
  {Herranz}, {Hildebrandt}, {Hivon}, {Huang}, {Jaffe}, {Jones}, {Karakci},
  {Keih{\"a}nen}, {Keskitalo}, {Kiiveri}, {Kim}, {Kisner}, {Knox},
  {Krachmalnicoff}, {Kunz}, {Kurki-Suonio}, {Lagache}, {Lamarre}, {Lasenby},
  {Lattanzi}, {Lawrence}, {Le Jeune}, {Lemos}, {Lesgourgues}, {Levrier},
  {Lewis}, {Liguori}, {Lilje}, {Lilley}, {Lindholm}, {L{\'o}pez-Caniego},
  {Lubin}, {Ma}, {Mac{\'\i}as-P{\'e}rez}, {Maggio}, {Maino}, {Mandolesi},
  {Mangilli}, {Marcos-Caballero}, {Maris}, {Martin}, {Martinelli},
  {Mart{\'\i}nez-Gonz{\'a}lez}, {Matarrese}, {Mauri}, {McEwen}, {Meinhold},
  {Melchiorri}, {Mennella}, {Migliaccio}, {Millea}, {Mitra},
  {Miville-Desch{\^e}nes}, {Molinari}, {Montier}, {Morgante}, {Moss}, {Natoli},
  {N{\o}rgaard-Nielsen}, {Pagano}, {Paoletti}, {Partridge}, {Patanchon},
  {Peiris}, {Perrotta}, {Pettorino}, {Piacentini}, {Polastri}, {Polenta},
  {Puget}, {Rachen}, {Reinecke}, {Remazeilles}, {Renzi}, {Rocha}, {Rosset},
  {Roudier}, {Rubi{\~n}o-Mart{\'\i}n}, {Ruiz-Granados}, {Salvati}, {Sandri},
  {Savelainen}, {Scott}, {Shellard}, {Sirignano}, {Sirri}, {Spencer},
  {Sunyaev}, {Suur-Uski}, {Tauber}, {Tavagnacco}, {Tenti}, {Toffolatti},
  {Tomasi}, {Trombetti}, {Valenziano}, {Valiviita}, {Van Tent}, {Vibert},
  {Vielva}, {Villa}, {Vittorio}, {Wandelt}, {Wehus}, {White}, {White},
  {Zacchei}, \& {Zonca}}]{Planck18}
{Planck Collaboration}, {Aghanim}, N., {Akrami}, Y., {et~al.} 2020, \aap, 641,
  A6, \dodoi{10.1051/0004-6361/201833910}

\bibitem[{{Plavin} {et~al.}(2022){Plavin}, {Paragi}, {Marcote}, {Keimpema},
  {Hessels}, {Nimmo}, {Vedantham}, \& {Spitler}}]{Plavin+22}
{Plavin}, A., {Paragi}, Z., {Marcote}, B., {et~al.} 2022, \mnras, 511, 6033,
  \dodoi{10.1093/mnras/stac500}

\bibitem[{{Plavin} {et~al.}(2019){Plavin}, {Kovalev}, {Pushkarev}, \&
  {Lobanov}}]{Plavin+19}
{Plavin}, A.~V., {Kovalev}, Y.~Y., {Pushkarev}, A.~B., \& {Lobanov}, A.~P.
  2019, \mnras, 485, 1822, \dodoi{10.1093/mnras/stz504}

\bibitem[{{Radcliffe} {et~al.}(2018){Radcliffe}, {Garrett}, {Muxlow},
  {Beswick}, {Barthel}, {Deller}, {Keimpema}, {Campbell}, \&
  {Wrigley}}]{Radcliffe+18}
{Radcliffe}, J.~F., {Garrett}, M.~A., {Muxlow}, T.~W.~B., {et~al.} 2018, \aap,
  619, A48, \dodoi{10.1051/0004-6361/201833399}

\bibitem[{{Ranasinghe} \& {Leahy}(2023)}]{Ranasinghe+2023}
{Ranasinghe}, S., \& {Leahy}, D. 2023, \apjs, 265, 53,
  \dodoi{10.3847/1538-4365/acc1de}

\bibitem[{{Ravi} {et~al.}(2022){Ravi}, {Law}, {Li}, {Aggarwal}, {Bhardwaj},
  {Burke-Spolaor}, {Connor}, {Lazio}, {Simard}, {Somalwar}, \&
  {Tendulkar}}]{RaviR67+22}
{Ravi}, V., {Law}, C.~J., {Li}, D., {et~al.} 2022, \mnras, 513, 982,
  \dodoi{10.1093/mnras/stac465}

\bibitem[{{Reines} {et~al.}(2020){Reines}, {Condon}, {Darling}, \&
  {Greene}}]{Reines2020}
{Reines}, A.~E., {Condon}, J.~J., {Darling}, J., \& {Greene}, J.~E. 2020, \apj,
  888, 36, \dodoi{10.3847/1538-4357/ab4999}

\bibitem[{{Resmi} {et~al.}(2021){Resmi}, {Vink}, \&
  {Ishwara-Chandra}}]{Resmi+21}
{Resmi}, L., {Vink}, J., \& {Ishwara-Chandra}, C.~H. 2021, \aap, 655, A102,
  \dodoi{10.1051/0004-6361/202039771}

\bibitem[{{Reynolds} {et~al.}(2012){Reynolds}, {Gaensler}, \&
  {Bocchino}}]{Reynolds+12}
{Reynolds}, S.~P., {Gaensler}, B.~M., \& {Bocchino}, F. 2012, \ssr, 166, 231,
  \dodoi{10.1007/s11214-011-9775-y}

\bibitem[{{Rhodes} {et~al.}(2023){Rhodes}, {Caleb}, {Stappers}, {Andersson},
  {Bezuidenhout}, {Driessen}, \& {Heywood}}]{Rhodes+23}
{Rhodes}, L., {Caleb}, M., {Stappers}, B.~W., {et~al.} 2023, arXiv e-prints,
  arXiv:2308.04298, \dodoi{10.48550/arXiv.2308.04298}

\bibitem[{{Robitaille} \& {Bressert}(2012)}]{aplpy}
{Robitaille}, T., \& {Bressert}, E. 2012, {APLpy: Astronomical Plotting Library
  in Python}, Astrophysics Source Code Library, record ascl:1208.017.
\newblock \doeprint{1208.017}

\bibitem[{Robitaille {et~al.}(2013)Robitaille, Tollerud, Greenfield,
  Droettboom, Bray, Aldcroft, Davis, Ginsburg, Price-Whelan, Kerzendorf,
  Conley, Crighton, Barbary, Muna, Ferguson, Grollier, Parikh, Nair,
  G{\"{u}}nther, Deil, Woillez, Conseil, Kramer, Turner, Singer, Fox, Weaver,
  Zabalza, Edwards, Azalee~Bostroem, Burke, Casey, Crawford, Dencheva, Ely,
  Jenness, Labrie, Lim, Pierfederici, Pontzen, Ptak, Refsdal, Servillat, \&
  Streicher}]{astropy1}
Robitaille, T.~P., Tollerud, E.~J., Greenfield, P., {et~al.} 2013, Astronomy
  {\&} Astrophysics, 558, 9, \dodoi{10.1051/0004-6361/201322068}

\bibitem[{{Sargent} {et~al.}(2022){Sargent}, {Johnson}, {Reines}, {Secrest},
  {van der Horst}, {Cigan}, {Darling}, \& {Greene}}]{Sargent2022}
{Sargent}, A.~J., {Johnson}, M.~C., {Reines}, A.~E., {et~al.} 2022, \apj, 933,
  160, \dodoi{10.3847/1538-4357/ac74be}

\bibitem[{{Shepherd} {et~al.}(1994){Shepherd}, {Pearson}, \& {Taylor}}]{DIFMAP}
{Shepherd}, M.~C., {Pearson}, T.~J., \& {Taylor}, G.~B. 1994, in Bulletin of
  the American Astronomical Society, Vol.~26, 987--989

\bibitem[{{Spitler} {et~al.}(2016){Spitler}, {Scholz}, {Hessels}, {Bogdanov},
  {Brazier}, {Camilo}, {Chatterjee}, {Cordes}, {Crawford}, {Deneva}, {Ferdman},
  {Freire}, {Kaspi}, {Lazarus}, {Lynch}, {Madsen}, {McLaughlin}, {Patel},
  {Ransom}, {Seymour}, {Stairs}, {Stappers}, {van Leeuwen}, \&
  {Zhu}}]{Spitler+16}
{Spitler}, L.~G., {Scholz}, P., {Hessels}, J.~W.~T., {et~al.} 2016, \nat, 531,
  202, \dodoi{10.1038/nature17168}

\bibitem[{{Sridhar} \& {Metzger}(2022)}]{Sridhar&Metzger22}
{Sridhar}, N., \& {Metzger}, B.~D. 2022, \apj, 937, 5,
  \dodoi{10.3847/1538-4357/ac8a4a}

\bibitem[{{Sridhar} {et~al.}(2021){Sridhar}, {Metzger}, {Beniamini},
  {Margalit}, {Renzo}, {Sironi}, \& {Kovlakas}}]{Sridhar+21b}
{Sridhar}, N., {Metzger}, B.~D., {Beniamini}, P., {et~al.} 2021, \apj, 917, 13,
  \dodoi{10.3847/1538-4357/ac0140}

\bibitem[{{Sridhar} {et~al.}(2022){Sridhar}, {Metzger}, \&
  {Fang}}]{Sridhar+23b}
{Sridhar}, N., {Metzger}, B.~D., \& {Fang}, K. 2022, arXiv e-prints,
  arXiv:2212.11236.
\newblock \doarXiv{2212.11236}

\bibitem[{{Tendulkar} {et~al.}(2017){Tendulkar}, {Bassa}, {Cordes}, {Bower},
  {Law}, {Chatterjee}, {Adams}, {Bogdanov}, {Burke-Spolaor}, {Butler},
  {Demorest}, {Hessels}, {Kaspi}, {Lazio}, {Maddox}, {Marcote}, {McLaughlin},
  {Paragi}, {Ransom}, {Scholz}, {Seymour}, {Spitler}, {van Langevelde}, \&
  {Wharton}}]{Tendulkar17}
{Tendulkar}, S.~P., {Bassa}, C.~G., {Cordes}, J.~M., {et~al.} 2017, \apjl, 834,
  L7, \dodoi{10.3847/2041-8213/834/2/L7}

\bibitem[{{Uro{\v{s}}evi{\'c}} {et~al.}(2005){Uro{\v{s}}evi{\'c}}, {Pannuti},
  {Duric}, \& {Theodorou}}]{eSN}
{Uro{\v{s}}evi{\'c}}, D., {Pannuti}, T.~G., {Duric}, N., \& {Theodorou}, A.
  2005, \aap, 435, 437, \dodoi{10.1051/0004-6361:20042535}

\bibitem[{van Bemmel {et~al.}(2022)van Bemmel, Kettenis, Small, Janssen,
  Moellenbrock, Petry, Goddi, Linford, Rygl, Liuzzo, Marcote, Bayandina,
  Schweighart, Verkouter, Keimpema, Szomoru, \& van Langevelde}]{Bemmel_2022}
van Bemmel, I.~M., Kettenis, M., Small, D., {et~al.} 2022, Publications of the
  Astronomical Society of the Pacific, 134, 114502,
  \dodoi{10.1088/1538-3873/ac81ed}

\bibitem[{{van Straten} \& {Bailes}(2011{\natexlab{a}})}]{van_Straten+11}
{van Straten}, W., \& {Bailes}, M. 2011{\natexlab{a}}, \pasa, 28, 1,
  \dodoi{10.1071/AS10021}

\bibitem[{{van Straten} \& {Bailes}(2011{\natexlab{b}})}]{vanstraten_2011_pasa}
---. 2011{\natexlab{b}}, \pasa, 28, 1, \dodoi{10.1071/AS10021}

\bibitem[{{van Straten} {et~al.}(2012){van Straten}, {Demorest}, \&
  {Oslowski}}]{vanstraten_2012_art}
{van Straten}, W., {Demorest}, P., \& {Oslowski}, S. 2012, Astronomical
  Research and Technology, 9, 237.
\newblock \doarXiv{1205.6276}

\bibitem[{{Vohl} {et~al.}(2023){Vohl}, {Vedantham}, {Hessels}, {Bassa}, {Cook},
  {Kaplan}, {Shimwell}, \& {Zhang}}]{Vohl2023}
{Vohl}, D., {Vedantham}, H.~K., {Hessels}, J.~W.~T., {et~al.} 2023, arXiv
  e-prints, arXiv:2303.12598, \dodoi{10.48550/arXiv.2303.12598}

\bibitem[{{Waxman}(2017)}]{Waxman+17}
{Waxman}, E. 2017, \apj, 842, 34, \dodoi{10.3847/1538-4357/aa713e}

\bibitem[{{Yang} {et~al.}(2020){Yang}, {Li}, \& {Zhang}}]{Yang+20}
{Yang}, Y.-P., {Li}, Q.-C., \& {Zhang}, B. 2020, \apj, 895, 7,
  \dodoi{10.3847/1538-4357/ab88ab}

\bibitem[{{Yao} {et~al.}(2017){Yao}, {Manchester}, \& {Wang}}]{YMW16}
{Yao}, J.~M., {Manchester}, R.~N., \& {Wang}, N. 2017, \apj, 835, 29,
  \dodoi{10.3847/1538-4357/835/1/29}

\bibitem[{{Zhang}(2018)}]{Zhang+18}
{Zhang}, B. 2018, \apjl, 854, L21, \dodoi{10.3847/2041-8213/aaadba}

\bibitem[{{Zhang} {et~al.}(2023){Zhang}, {Yu}, {Law}, {Li}, {Chatterjee},
  {Demorest}, {Yan}, {Niu}, {Aggarwal}, {Anna-Thomas}, {Burke-Spolaor},
  {Connor}, {Tsai}, {Zhu}, \& {Luo}}]{Zhang+23}
{Zhang}, X., {Yu}, W., {Law}, C., {et~al.} 2023, arXiv e-prints,
  arXiv:2307.16355, \dodoi{10.48550/arXiv.2307.16355}

\end{thebibliography}
\bibliographystyle{aasjournal}
\end{document}